\documentclass[pdflatex,sn-mathphys-num,iicol]{sn-jnl}% Math and Physical Sciences Numbered Reference Style
\usepackage{graphicx}%
\usepackage{multirow}%
\usepackage{amsmath,amssymb,amsfonts}%
\usepackage{amsthm}%
\usepackage{mathrsfs}%
\usepackage[title]{appendix}%
\usepackage{xcolor}%
\usepackage{textcomp}%
\usepackage{manyfoot}%
\usepackage{booktabs}%
\usepackage{algorithm}%
\usepackage{algorithmicx}%
\usepackage{algpseudocode}%
\usepackage{listings}%
\usepackage{subcaption}%
\usepackage{multirow}
\usepackage{stfloats}
\usepackage{fancyhdr}

\fancypagestyle{reportfirstpage}{
    \fancyhf{}
    \fancyhead[R]{\color{gray} MCNET-26-24}
    
}
\usepackage{todonotes}
\presetkeys{todonotes}{inline}{}
\usepackage{amsmath}
\usepackage{amssymb}
\usepackage{array}
\usepackage{calc}
\usepackage{longtable}
\usepackage{multirow}
\usepackage{tensor}
\usepackage{upgreek}
\usepackage{graphicx}
\graphicspath{{figures/}}
\usepackage{xspace}
\usepackage{listings}
\usepackage[section]{placeins}
\usepackage{microtype}
\usepackage{
  pgf,
  tikz}
\usetikzlibrary{
  patterns,
  shapes.multipart,
  arrows,
  trees,
  scopes,
  decorations.pathreplacing,
  decorations.pathmorphing,
  decorations.markings,
  decorations.text,
  positioning,
  calc
}

\usepackage[utf8]{inputenc}
\usepackage{mciteplus}
\newcommand{\Pepper}{P\protect\scalebox{0.8}{EPPER}\xspace}

\newcommand{\Amegic}{A\protect\scalebox{0.8}{MEGIC}\xspace}
\newcommand{\Comix}{C\protect\scalebox{0.8}{OMIX}\xspace}
\newcommand{\Chili}{C\protect\scalebox{0.8}{HILI}\xspace}
\newcommand{\Whizard}{W\protect\scalebox{0.8}{HIZARD}\xspace}
\newcommand{\Vegas}{V\protect\scalebox{0.8}{EGAS}\xspace}

\newcommand{\MadGraph}{M\protect\scalebox{0.8}{AD}G\protect\scalebox{0.8}{RAPH}\xspace}

\newcommand{\pytorch}{\textsc{PyTorch}\xspace}
\newcommand{\torchdyn}{\textsc{TorchDyn}\xspace}
\theoremstyle{thmstyleone}%
\theoremstyle{thmstyletwo}%

\theoremstyle{thmstylethree}%

\newcommand{\diff}{\mathop{}\!\mathrm{d}}
\makeatletter
\newcommand{\spx}[1]{%
  \if\relax\detokenize{#1}\relax
    \expandafter\@gobble
  \else
    \expandafter\@firstofone
  \fi
  {^{#1}}%
}
\makeatother
\newcommand\pd[3][]{\frac{\partial\spx{#1}#2}{\partial#3\spx{#1}}}
\newcommand{\od}[3][]{\frac{\diff\spx{#1}#2}{\diff#3\spx{#1}}}

\newsavebox{\firstvecbox}
\newsavebox{\secondvecbox}
\newsavebox{\tempvecbox}

\newcommand{\topalignedbmatrix}[1]{%
  \sbox{\tempvecbox}{$\begin{bmatrix}#1\end{bmatrix}$}%
  \raisebox{\dimexpr\ht\firstvecbox-\ht\tempvecbox\relax}{\usebox{\tempvecbox}}%
}
\sbox{\firstvecbox}{$
\begin{bmatrix}
    z_N^{(1)} \\
    z_N^{(2)} \\
    \vdots \\
    z_N^{(d_{N})}
\end{bmatrix}
$}

\sbox{\secondvecbox}{$
\begin{bmatrix}
    c_{1}\\
    c_{2}\\
    \vdots \\
    c_{n_{\text{hel}}}\\
\end{bmatrix}
$}

\begin{document}

\title{Too good to go: Upcycling Phase-Space Points for Multijet Processes}

\author[1]{\fnm{Konrad} \sur{Helms}}\email{konrad.helms@theorie.physik.uni-goettingen.de}

\author[1]{\fnm{Timo} \sur{Jan\ss{}en}}\email{timo.janssen@theorie.physik.uni-goettingen.de}

\author[1]{\fnm{Steffen} \sur{Schumann}}\email{steffen.schumann@phys.uni-goettingen.de}

\affil[1]{\orgdiv{Institute for Theoretical Physics}, \orgname{Georg-August-University G\"ottingen}, \orgaddress{\street{Friedrich-Hund-Platz 1}, \city{G\"ottingen}, \postcode{37077}, \country{Germany}}}

\abstract{\unboldmath
  The efficient sampling of high-dimensional phase spaces is a major challenge for Monte Carlo event generators, 
  as for high-multiplicity final states the evaluation of scattering matrix elements becomes computationally expensive. 
  We here introduce a training strategy that significantly reduces the cost of adapting the samplers, while 
  delivering samplers that outperform the current benchmarks.
  The method exploits the nested structure of phase spaces, where an $(N+1)$-particle phase space 
  factorises into an $N$-particle and a one-particle phase space. We thereby assume that an efficient sampler for the 
  corresponding $N$-particle phase space is already available, as is the case in stacks of QCD $X+n$-jets processes. 
  By augmenting an $N$-particle to an $(N+1)$-particle dataset, we obtain a training sample that more closely resembles the 
  integrand and is statistically larger than, for example, a uniformly sampled one. Starting the adaptation phase of the 
  sampler with a well-sampled $N$-particle core accelerates learning of the full $(N+1)$-particle phase-space density. 
  Since the augmented sample is only used as an initial proposal distribution, unbiased Monte Carlo estimates are still 
  guaranteed by exact event weighting. The approach is agnostic to the trained sampler and can be applied to 
  machine-learning-based methods as well as more traditional algorithms such as \Vegas. We demonstrate the 
  reduction in matrix-element evaluations and final performance increase for jet-associated Drell--Yan and
  top-pair production at the LHC, using Continuous Normalising Flow samplers.
}




\maketitle
\thispagestyle{reportfirstpage}

\section{Introduction}
\label{sec:introduction}
Monte Carlo event generators play a central role in the analysis and interpretation
of high-energy scattering experiments, such as those performed at the Large Hadron Collider
(LHC). They provide detailed descriptions of individual scattering events based on a factorised
event-evolution approach~\cite{Buckley:2011ms,Campbell:2022qmc}. Starting from a partonic
hard-process flavour and momentum configuration, they address subsequent QCD brems\-strahlung
through parton-shower simulations, and invoke phenomenological models to account for the
underlying event and the transition of partons to hadrons.

In particular for complex hard processes, i.e.\ high-multiplicity final states,
the efficient exploration of the associated high-dimensional phase space presents a severe
challenge. Suboptimal sampling thereby results in wide-spread event weights, corresponding
to significant variances for the corresponding cross-section estimates, and typically rather
low event-unweighting efficiencies, see for example~\cite{Hoche:2019flt}. The mitigation of
this problem is the arena of variance-reduction techniques, in particular importance-sampling
and stratification methods, that aim to adapt the distribution of random samples as close as
possible to the desired target distribution. To this end, specialised matrix-element generators
such as \Amegic~\cite{Krauss:2001iv}, \Comix~\cite{Gleisberg:2008fv}, \MadGraph~\cite{Maltoni:2002qb}
or \Whizard~\cite{Kilian:2007gr} construct process-specific adaptable phase-space maps which
reflect the topologies, propagator virtualities, and angular-variable dependences of
individual partonic channels. 

Given such phase-space map, its corresponding sampling density can be further optimised by
redistributing its uniform input variables. In existing Monte Carlo event generators
such remapping is typically handled by the \Vegas algorithm~\cite{Lepage:1977sw,Ohl:1998jn}.
More recently, modern machine-learning algorithms, and in particular Normalising
Flows~\cite{tabak:2010,Tabak:2013cnz,pmlr-v37-rezende15,papamakarios:2021},
have been instrumented for this task and significant performance gains can be
realised~\cite{Bothmann:2020ywa,Gao:2020zvv,Heimel:2022wyj,Verheyen:2022tov,Heimel:2023ngj,
  Deutschmann:2024lml,Kofler:2024efb,ElBaz:2025qjp,Janssen:2025zke,Bothmann:2025lwg,DeCrescenzo:2026tsp,Bothmann:2026dar,Heimel:2026cxh}.
The training of the sampler, being it the adaptation of \Vegas grids or the adjustment of the parameters
of the Normalising Flow, is typically done independently for distinct processes. However, in
particular for complex flow models training is computationally intense and requires a non-negligible
number of samples from the typically expensive target function, i.e.\ the fully differential scattering
matrix element. This can partially be ameliorated through iterative-training procedures, where
after an initial adaptation phase based on uniform samples, new data points get generated
by the sampler trained in the previous iteration, see for example~\cite{Bothmann:2025lwg,Janssen:2025zke,akhoundsadegh2024}. 

A particular challenge in the description of LHC events is the simulation of multijet processes,
i.e.\ the production of a given core process in association with a variable number of QCD jets.
Examples for relevant core processes include vector-boson, top-quark, (di-)photon, or even
Higgs-boson production. All these signatures are interesting in their own right, as they provide
powerful tests of perturbative QCD. They can, for example, be used to infer about scaling
patterns of jet-production rates~\cite{Gerwick:2011tm,Gerwick:2012hq}, to extract the strong-coupling
constant~\cite{Johnson:2017ttl}, or to determine parton-density functions~\cite{Bailey:2020ooq,NNPDF:2021njg}.
However, most prominently, they form omnipresent backgrounds that contribute in almost every analysis,
e.g.\ when searching for physics beyond the Standard Model. A particular challenge for modelling
multijet-associated production modes is the presence of hierarchical scales, including the momentum
transfer in the hard process, the various jet transverse momenta, and the actual jet-resolution scale,
what can give rise to sizeable higher-order corrections, see for example~\cite{Rubin:2010xp}. To
adequately model these variable jet-multiplicity processes pure parton-shower simulations are often
insufficient. Instead, for this purpose dedicated matrix-element plus parton-shower merging schemes
have been developed, see for example
Refs.~\cite{Catani:2001cc,Lonnblad:2001iq,Krauss:2002up,Lavesson:2005xu,Hoeche:2009rj,Andersen:2011hs}
and Ref.~\cite{Alwall:2007fs} for a comparative study. Largely independent of the details of the approach
used, samples drawn from a stack of hard processes with increasing jet multiplicity are needed, which
are separated by a jet-resolution criterion, often dubbed the \emph{merging scale}. To this end,
samplers for each contributing multiplicity need to be trained. Conventionally, this is done in a 
fully independent manner. In particular for the high-multiplicity contributions this is computationally
expensive.

We here propose a new adaptation strategy for the samplers of a stack of multijet processes which largely
reduces the computational effort to train the integrator for high-multiplicity processes. Based on a
factorised phase-space mapping supplemented with a Normalising Flow remapping, we utilise the optimised
sampler for a given jet multiplicity, augment it by the additional dimensions of an extra particle, and
use this to generate training data for the initial stage of an iterative training for the next multiplicity
process. This constitutes a form of cross-multiplicity knowledge transfer: rather than training each sampler entirely
independently, the approximate distribution and transport learned at multiplicity $n$ are reused to bootstrap training
at multiplicity $n+1$. For our study we employ the \Chili~\cite{Bothmann:2023siu} phase-space parameterisation, remap
its inputs with Continuous Normalising Flows (CNFs)~\cite{Chen:2018}, and use Conditional
Flow Matching (CFM)~\cite{Lipman:2023,Liu:2022,Albergo:2023building,Albergo:2023stochastic,Tong:2024,Pooladian:2023}
for their training. To illustrate the potential of our new training strategy, we apply it to the stack
of $d\bar{d}\to e^+e^-+n$-gluons and $gg\to t\bar{t}+n$-gluons processes with $n\leq 5$, respectively,
considering proton--proton collisions at $\sqrt{s}=13\,\text{TeV}$. The corresponding tree-level matrix elements we
obtain from the \Pepper generator~\cite{Bothmann:2023gew}. 

Our paper is organised as follows: In Section~\ref{sec:review} we briefly review the \Chili phase-space
parameterisation for jet-associated production processes, introduce the remapping approach for phase-space
variables, and provide a focused introduction to CNFs and CFM.
In Section~\ref{sec:knowledge-transfer} we introduce our phase-space augmentation technique and detail
the strategies used to successively train a stack of multijet processes. In Section~\ref{sec:application_to_multijet_production}
we present and discuss our results for Drell--Yan and top-quark pair production in association with gluons.
Finally, in Section~\ref{sec:conclusions} we conclude our study, discuss possible generalisations, and identify
interesting future research avenues. 

\section{Setting the Scene}
\label{sec:review}

The sampling of hadronic scattering events amounts to the integration of the hadronic
cross section for the production of a given $N$-particle final state $X_N$. The corresponding
cross-section integral for the process $pp\to X_N$ reads
\begin{equation}\label{eq:hadroic-cross-section}
  \begin{split}
    \sigma_{p p \to X_N} &= \sum_{a,b} \int_0^1 \diff x_a \int_0^1 \diff x_b\\
                         &\quad {}\times f^p_a(x_a, \mu_F) \ f^p_b(x_b, \mu_F)\\
                         &\quad {}\times\hat{\sigma}_{ab \to X_N}(x_a, x_b, \mu_R, \mu_F)\,,
  \end{split}
\end{equation}
with $f^p_{a/b}(x_{a/b},\mu_F)$ the density functions for parton flavours $a/b$ in the proton, respectively,
evaluated at factorisation scale $\mu_F$. The partonic cross section for the reaction $ab\to X_N$ is thereby given by
\begin{equation}
  \hat{\sigma}_{ab \to X_N}(\hat{s}) = \frac{1}{2\hat{s}} \int \diff \mathrm{\Phi}_N |\mathcal{M}_{ab \to X_N}|^2 \,,
  \label{eq:partonic-cross-section}
\end{equation}
with the partonic squared centre-of-mass energy 
\begin{equation}
  \hat{s} = (p_a  + p_b)^2 = x_ax_bs\,,
  \label{eq:partonic-com}
\end{equation}
and $|\mathcal{M}_{ab \to X_N}|^2$ the squared matrix element. The 
Lorentz invariant phase-space element $\diff \mathrm{\Phi}_N$ is given by
\begin{equation}
  \diff \mathrm{\Phi}_N = \left[\prod\limits_{i=1}^N\frac{d^3p_i}{(2\pi)^32E_i}\right](2\pi)^4\delta^{(4)}\left(p_{ab}-\sum\limits_{i=1}^Np_i\right)\,,
  \label{eq:LIPS}
\end{equation}
with $p_{ab} = p_a + p_b$ and $E_i=\sqrt{\vec{p}^2_i+m_i^2}$. The phase-space volume will typically be subject
to cuts, for example to regularise soft and collinear singularities, or to more efficiently
address a specific fiducial volume. Given the complexity of the integrand in particular for
high-multiplicity final states, dedicated phase-space maps, supplemented by suitable probability densities for
the integration variables, need to be employed to achieve reasonable performance. In particular multichannel
techniques are widely used, which map out the diagrammatic topologies contributing to the integrand,
see for example~\cite{Kleiss:1994qy,Krauss:2001iv,Maltoni:2002qb,Gleisberg:2008fv}. In this work we rather
focus on approaches which employ the factorisation properties of the $N$-particle phase-space
measure~\cite{Byckling:1969luw,Byckling:1969sx}. Such maps can potentially be constructed along with the
recursion used for constructing the matrix-element expressions~\cite{Gleisberg:2008fv}. For our work,
however, we make use of the fully factorised \Chili integrator presented in Ref.~\cite{Bothmann:2023siu},
that, by construction, targets multijet-production processes. It relies on a simple parameterisation of a
particle's phase space in terms of its squared transverse momentum ($\diff p^2_\perp$), rapidity ($\diff y$),
and azimuthal angle ($\diff \varphi$). This choice of variables allows one to directly account for typical
phase-space restrictions in multijet processes, namely cuts on the jets' transverse momenta and rapidities.
Most important for our application, the \Chili $(N+1)$-particle phase space can be factorised into an 
$N$-particle phase space and a one-particle phase space, i.e.\
\begin{equation}
  \diff x_a \diff x_b \diff\mathrm{\Phi}_{N+1} \sim [\diff x_a \diff x_b \diff\mathrm{\Phi}_{N}] \times
  \diff\mathrm{\Phi}_{1}\,,
  \label{eq:phi_n_plus_one_phi_one}
\end{equation}
where, in deviation from Eq.~\eqref{eq:LIPS}, we use $\diff\mathrm{\Phi}_1$ to denote the three-dimensional
single-particle Lorentz-invariant measure.

In what follows, we will briefly review the basics of the \Chili phase-space integrator
and detail the specifics for our two applications, namely jet-associated Drell--Yan and
top-quark pair production. We will then provide a focused introduction to variable
remappings via CNFs and their training. 

\subsection{The \Chili Phase-Space Parameterisation}
\label{subsec:PS-parametrization}
The \Chili phase-space parameterisation realises a mapping $T_N : [0,1]^{d_N} \to \{x_a, x_b, p_1, \ldots, p_N\}$ of the
$d_{N}=3N-4+2$ dimensional unit hypercube
to the physical particle momenta. For final states exclusively containing QCD particles, the phase space is
parameterised in terms of the particles' lab-frame squared transverse momenta $p_{i, \perp}^2$, rapidities $y_i$, and azimuthal
angles $\varphi_i$. 
All final-state particles but one are parameterised in this way. To implement four-momentum conservation, one particle is
designated as the recoil, of which only the rapidity $y_{\text{rec}}$ is independent; its remaining degrees of freedom
are fixed by transverse-momentum balance against all other final-state particles, and its integration is performed in
combination with the light-cone momentum fractions of the initial-state particles $a$ and $b$. The recoil is a fixed
particle, chosen independently of the jet multiplicity: in what follows it is the $Z$ boson for $d\bar{d} \to e^+ e^- +
ng$ and the top quark for $gg \to t\bar{t} + ng$. In particular, all gluons are fully resolved at every multiplicity.
The set of coordinates
\begin{multline}\label{eq:chili_coordinates}
  \boldsymbol{z}^{\text{sampling}}_N =\\
  {} (p_{1,\perp}^2, y_1, \varphi_1, \ldots, p_{N-1,\perp}^2, y_{N-1}, \varphi_{N-1}, y_{\text{rec}})
\end{multline}
fully determines the initial- and final-state momenta, where for top-pair production particles $1,\ldots,n$ are the
gluons and particle $N-1$ is the $\bar{t}$.  
Accordingly, the $gg \to t\bar{t} + ng$ phase space, consisting of
$N=n+2$ final-state particles, is pa\-ra\-me\-te\-ri\-sed as 
\begin{multline}\label{eq:qcd-ps}
    \diff x_a \diff x_b \diff \mathrm{\Phi}_N(a,b;1,\dots,N) =\\ 
    {} \frac{2\pi}{s}\left[\prod_{\substack{i=1 \\ i \neq \text{rec}}}^{N}\frac{1}{(4\pi)^2} \diff p_{i, \perp}^2  \diff y_i \frac{\diff \varphi_i}{2\pi}\right]\diff y_{\text{rec}}\,\,.
\end{multline}
This expression defines the (constant) kinematic Jacobian $\mathcal{J}_N^{\text{kin}}$.

The coordinates are generated in the order given in Eq.~\eqref{eq:chili_coordinates}. For the remainder of this work it is
convenient to reorder them by a fixed permutation $P_N$,
\begin{multline}
  \boldsymbol{z}_N = P_N \boldsymbol{z}^{\text{sampling}}_N =\\
  {} \big(p_{\bar{t}, \perp}^2, y_{\bar{t}}, \varphi_{\bar{t}}, y_{\text{rec}}, p^2_{1,\perp}, y_1, \varphi_1, \dots, p^2_{n\perp}, y_n, \varphi_n\big)\,,
\end{multline}
and correspondingly for Drell--Yan production, so that the four coordinates not associated with a resolved gluon come
first and increasing the gluon multiplicity appends the new gluon's coordinates at the end of the vector. Since the same
four coordinates are moved to the front at every multiplicity, appending the new gluon's coordinates in this ordering is
equivalent to inserting them at position $3n+1$ of the sampling-order vector, as done in the implementation.

The coordinates $\boldsymbol{z}_N$ are generated from uniform variables
\begin{equation}
  \boldsymbol{u}_N = (u_1, \ldots, u_{d_N}) \sim \mathcal{U}([0, 1]^{d_N}) \,.
\end{equation}
The variables $y_i$, $y_{\text{rec}}$ and $\varphi_i$ are generated uniformly, i.e.\
\begin{align}
  y_i &= y_\text{min}+u_{y_i} \,(y_\text{max}-y_\text{min})\,,\label{eq:y_map}\\
  \varphi_i &= 2\pi\,u_{\varphi_i}\,,\label{eq:phi_map}
\end{align}
with $u_{y_i}, u_{\varphi_i} \sim {\cal{U}}(0,1)$. For all particles subject to a transverse-momentum cut, i.e.\ the
gluons, the transverse momenta are generated by
\begin{equation}\label{eq:pT_map_g}
  p_{i,\perp}^2 = \Biggl[ \frac{1-u_{p_{i,\perp}^2}}{p_{\perp,\text{min}}^2} +
  \frac{u_{p_{i,\perp}^2}}{p_{\perp,\text{max}}^2} \Biggr]^{-1}
\end{equation}
with $u_{p_{i,\perp}^2} \sim {\cal{U}}(0,1)$, conforming to a density ${\cal{P}}(p_\perp^2) \propto (p_\perp^2)^{-2}$.
The lower limit, $p_{\perp,\text{min}}^2$, corresponds to the jet transverse-momentum cut and the
upper limit is given by $p_{\perp,\text{max}}^2 = s/4$. The massive $\bar{t}$, whose effective
threshold is set by its mass rather than by the jet cut, is instead generated according to
\begin{equation}\label{eq:pT_dist_top}
  {\cal{P}}(p_{\bar{t},\perp}) \propto \frac{1}{(2m_t + p_{\bar{t},\perp})^2}\,.
\end{equation}
The $\bar{t}$ is in addition balanced against the already-generated particles through shifting its transverse momentum
by $\bigl(\sum_{i=1}^n \boldsymbol{p}_{i,\perp}\bigr) / 2$, the remainder being absorbed by the recoil particle.

Combining the kinematic and sampling transformations gives
\begin{equation}\label{eq:chili_jacobian}
  \diff x_a \diff x_b \diff \Phi_N = \mathcal{J}_N^{\text{\Chili}}(\boldsymbol{u}_N) \diff^{d_N}\boldsymbol{u}_N\,,
\end{equation}
with
\begin{equation}
  \mathcal{J}_N^{\text{\Chili}}(\boldsymbol{u}_N) = \mathcal{J}_N^{\text{kin}} \biggl\lvert \det \pd{\boldsymbol{z}_N}{\boldsymbol{u}_N}
  \biggr\rvert \,.
\end{equation}
This construction defines the map
\begin{equation}
  T_N = K_N \circ S_N\,,
\end{equation}
where $S_N$ maps uniformly distributed variables to the \Chili coordinates and $K_N$ reconstructs the physical momenta.

The parameterisation for the second considered process, $d\bar{d} \to e^+e^- + ng$, is slightly different.
Here, the $N=n+2$ particle phase space is first decomposed into lower-multiplicity phase-space elements
\begin{equation}
  \begin{split}
    \diff\mathrm{\Phi}_N(a,b;1,\dots,N) &= \diff\mathrm{\Phi}_{n+1}(a,b;Z, 3, \dots, N)\\
                                        &\quad {}\times \frac{\diff s_{Z}}{2\pi} \times \diff\mathrm{\Phi}_2(Z;e^+,e^-)
  \end{split}
\end{equation}
and then, to respect four-momentum conservation, integrated along with the momentum fractions $x_a$ and $x_b$:
\begin{multline}
    \diff x_a \diff x_b \diff \mathrm{\Phi}_N(a,b;1,\dots,N) =  \\
    \frac{2\pi}{s}\left[\prod_{i=1}^{N-2}\frac{1}{(4\pi)^2} \diff p_{i, \perp}^2  \diff y_i \frac{\diff \varphi_i}{2\pi}\right] \frac{\diff s_{Z}}{2\pi} \\
    {}\times \diff y_{\text{rec}} \frac{1}{(4\pi)^2}\frac{\sqrt{(p_{e^+}p_{e^-})^2-p_{e^+}^2p_{e^-}^2}}{(p_{e^+}+p_{e^-})^2} \\
    {}\times \diff\cos\theta_1^{\{e^+,e^-\}}\diff\varphi_1^{\{e^+,e^-\}}\,.
\end{multline}
This way, the production of the $Z$ boson, with virtuality $s_{Z} = p_{Z}^2$, and its subsequent decay $Z\to e^+e^-$
is factorised from the QCD part. The virtuality thereby follows a Breit--Wigner distribution, while
$\cos\theta_1$ and $\varphi_1$, considered in the centre-of-mass frame of the combined momentum $p_Z=p_{e^+}+p_{e^-}$,
are distributed uniformly. The phase space $\mathrm{\Phi}_{n+1}$ of the $n$ gluons and the (off-shell) $Z$ is again
parameterised according to Eq.~\eqref{eq:qcd-ps}, with the $Z$ acting as the recoil particle. 

Using the above parameterisations, with a fixed ordering of the final-state particles and the same choice of recoil
particle at every multiplicity, the coordinate vectors are nested:
\begin{align}
  \boldsymbol{z}_{X+(n+1)g} &= (\boldsymbol{z}_{X+ng}, p_{n+1,\perp}^2, y_{n+1}, \varphi_{n+1})\,,\label{eq:coord_nesting}\\
  d_{n+1} &= d_n+3\,.
\end{align}
In the \Chili parameterisation, we can now identify the one-particle phase space in Eq.~\eqref{eq:phi_n_plus_one_phi_one},
with the additional particle's three kinematic degrees of freedom, and write the $(N+1)$-particle phase space as
\begin{multline}
  \diff x_a \diff x_b \diff\mathrm{\Phi}_{N+1} =\\
  {} [\diff x_a \diff x_b \diff \mathrm{\Phi}_{N}] \times \frac{1}{(4\pi)^2 2\pi} \diff p_{\perp}^2  \diff y \,\diff \varphi\,.
\end{multline}
This should be understood as a nesting of the coordinate parameterisations rather than a literal inclusion of physical
phase spaces. 
Since the recoil is taken by the same particle at every multiplicity and the gluons are generated in index order, the map from
the $3n$ gluon coordinates to the first $n$ gluon's momenta, and their coordinate ranges, are independent of
$n$. Adding a gluon affects only the remaining four coordinates and the reconstructed
momentum fractions $x_a$, $x_b$. The additional particle nonetheless draws energy from the collision, so a configuration
realisable at multiplicity $n$ may require $x_a$ or $x_b$ to exceed unity at multiplicity $n+1$, while
transverse-momentum cancellation can conversely render an $(n+1)$-particle point valid whose $n$-particle counterpart is
not. The two sets of valid points hence overlap substantially without either containing the other.

\subsection{Phase-space Sampling and Remapping}
\label{subsec:ps-sampling-background}
Through Eqs.~\eqref{eq:y_map}--\eqref{eq:pT_dist_top}, the \Chili map transforms a point $\boldsymbol{u}_N \in [0,1]^{d_N}$ into the
corresponding initial-state momentum fractions and final-state momenta. The differential cross section pulled back to
the unit hypercube is
\begin{equation}\label{eq:hypercube_integrand}
  \begin{split}
    \mathcal{F}_N(\boldsymbol{u}_N) &= \sum_{a,b} f_a^p(x_a, \mu_F) f_b^p(x_b, \mu_F) \\
                                    &\quad {}\times \frac{\lvert\mathcal{M}_{ab \to X_N}\rvert^2}{2\hat{s}} \mathcal{J}_N^{\text{\Chili}}(\boldsymbol{u}_N) \Theta_{\text{cuts}}\,,
  \end{split}
\end{equation}
where all kinematic quantities on the right-hand side are evaluated at $T_N(\boldsymbol{u}_N)$. The total cross section
and, more generally, a weighted cross section for an observable $O$ then read
\begin{align}
  \sigma_{pp \to X_N} &= \int_{[0,1]^{d_N}} \diff^{d_N} \boldsymbol{u}_N \mathcal{F}_N(\boldsymbol{u}_N)\,, \\
  \int O \diff \sigma &= \int_{[0,1]^{d_N}} \diff^{d_N} \boldsymbol{u}_N \mathcal{F}_N(\boldsymbol{u}_N) O
  \label{eq:observable_integral}
  \bigl(T_N(\boldsymbol{u}_N)\bigr)\,.
\end{align}

When considering importance sampling, points are drawn from a normalised density $q_N(\boldsymbol{u}_N)$ and assigned the event
weights
\begin{equation}\label{eq:event_weight}
  w_i = \frac{\mathcal{F}_N(\boldsymbol{u}_N^{(i)})}{q_N(\boldsymbol{u}_N^{(i)})}\,.
\end{equation}
A sample of $N_{\text{ev}}$ weighted events therefore yields the Monte Carlo estimates
\begin{align}
  \sigma &\simeq \frac{1}{N_{\text{ev}}} \sum_{i=1}^{N_{\text{ev}}} w_i, \label{eq:mc_estimate}\\
  \int O \diff \sigma &\simeq \frac{1}{N_{\text{ev}}} \sum_{i=1}^{N_{\text{ev}}} w_i O \bigl(T_N(\boldsymbol{u}_N^{(i)})\bigr)\,,
\end{align}
with variance
\begin{equation}
  \operatorname{Var}[\sigma] = \frac{1}{N_{\text{ev}}} \Biggl[ \int_{[0,1]^{d_N}} \diff^{d_N} \boldsymbol{u}_N
\frac{\mathcal{F}_N^2(\boldsymbol{u}_N)}{q_N(\boldsymbol{u}_N)} - \sigma^2 \Biggr]\,.
\end{equation}
The optimal importance density is proportional to the differential cross section, $q_N^\star = \mathcal{F}_N / \sigma$. In
this limit, all events have the same weight and the generated momenta directly follow the normalised differential
cross-section distribution. In practice, approximating this density reduces both the variance of the cross-section
estimate and the spread of event weights.

Uniform sampling of the \Chili variables corresponds to $q_N(\boldsymbol{u}_N) = 1$. Although the \Chili parameterisation
captures important features of jet-dominated processes, it generally provides only a crude approximation to the fully
differential cross section. The sampling density can be improved by composing the \Chili map $T_N$ with an invertible remapping of the unit hypercube,
\begin{equation}\label{eq:remap}
  \boldsymbol{u}_N = R_{\theta,N}(\boldsymbol{v}_N)\,,\quad \boldsymbol{v}_N \sim \mathcal{U}([0,1]^{d_N})\,,
\end{equation}
with learnable parameters $\theta$. The complete map from uniform random numbers to physical momenta is then
$T_N \circ R_{\theta,N}$, with Jacobian
\begin{equation}
  \mathcal{J}_N^{\text{full}}(\boldsymbol{v}_N) = \mathcal{J}_N^{\text{\Chili}}\bigl(R_{\theta,N}(\boldsymbol{v}_N)\bigr) \biggl\lvert \det
    \pd{R_{\theta,N}(\boldsymbol{v}_N)}{\boldsymbol{v}_N} \biggr\rvert\,.
\end{equation}
Accordingly, the event weight is obtained from Eq.~\eqref{eq:hypercube_integrand} by replacing $\mathcal{J}_N^{\text{\Chili}}$
with $\mathcal{J}_N^{\text{full}}$. The remapping $R_{\theta,N}$ may, for example, be realised by a \Vegas grid or a Normalising
Flow. In the following section, we consider CNFs~\cite{Chen:2018} trained using CFM~\cite{Lipman:2023,Tong:2024}.

The performance of the presented sampling methods is assessed using several standard metrics. 
Since the primary aim usually is to compute physical observables, such as the cross section
of a process, we track the statistical error of the MC estimate
\begin{equation}
    \sigma_{\text{error}} = \sqrt{\frac{\text{Var}(w)}{N_{\text{ev}}}}\,,
    \label{eq:integration_error}
\end{equation}
where $w$ are the MC weights, see Eq.~\eqref{eq:mc_estimate}, and $N_{\text{ev}}$ is the number of events.

Moreover, only a subset of the generated events satisfies the imposed phase-space cuts. The 
fraction of generated points that pass these cuts is referred to as phase-space efficiency.

Additionally, we compute the relative Kish effective sample size~\cite{Kish:1965} 
\begin{equation}
  \frac{N_{\text{eff}}}{N_{\text{ev}}} = \frac{\langle w\rangle^2}{\langle w^2\rangle}
    \label{eq:kish_ess}
\end{equation}
which measures the statistical power of the weighted sample. 
In the ideal case of a perfectly sampled dataset, all weights are identical. The 
effective sample size for such a dataset consequently equals one. Instead, when 
$\frac{N_{\text{eff}}}{N_{\text{ev}}} \ll 1 $, the dataset behaves like a statistically much smaller one.

In high energy physics, weighted event samples are often converted to independent 
and identically distributed samples by means of rejection sampling. The average acceptance
probability of events, i.e.\
\begin{equation}
    \epsilon = \frac{\langle w \rangle}{C}\,,
    \label{eq:uw_efficiency}
\end{equation}
defines the so-called unweighting efficiency, where $C = \text{max}(w(\boldsymbol{u}_N))$ is the 
largest possible MC weight that can occur. Given $C$ needs to be estimated from finite samples, 
the unweighting efficiency can be overly sensitive to rare outliers. To mitigate this, we replace 
$C$ by an effective maximal weight. This bound is chosen such that weights exceeding it contribute 
at most 0.1\%, or 1\%, to the total MC integral~\cite{Danziger:2021eeg}. The corresponding effective
unweighting efficiencies 
are denoted by  $\epsilon_{0.1\%}$ and $\epsilon_{1\%}$, respectively.

The primary goal of our training strategy is to lower the computational cost of training 
phase-space point generators, especially for the high-gluon multiplicities where 
evaluating the corresponding matrix element is expensive. We therefore use the number 
of training events that require a matrix-element evaluation as our cost metric.
Because the benchmark networks and the networks trained on augmented data share the same architecture, batch size, and
number of epochs per iteration, the cost of the network training itself is identical per iteration and is not included
in this metric. Since the augmented models typically require fewer iterations to converge, this omission is
conservative: it discards a saving in favour of the proposed method. Networks that are initially 
trained on an augmented dataset start at zero matrix-element cost, as all phase-space 
points they see have already been generated during the training of a lower-multiplicity model.
Consequently, their training cost only comes from the datasets generated during their iterative 
training.

\subsection{ODE Flows and their Training}
\label{subsec:ml-background}
A CNF defines an invertible transport between a base density $q_0$, usually a standard
normal,  and a learned density
$q_1$ on $\mathbb{R}^d$~\cite{Chen:2018}. Starting from $\boldsymbol{x}_0 \sim q_0$, a time-dependent vector field
$\boldsymbol{v}_{t,\theta} :
[0,1] \times \mathbb{R}^d \to \mathbb{R}^d$, with learnable parameters $\theta$, determines the trajectory
$\boldsymbol{x}_t$ through the ordinary differential equation (ODE)
\begin{equation}
  \od{\boldsymbol{x}_t}{t} = \boldsymbol{v}_{t,\theta}(\boldsymbol{x}_t), \qquad \boldsymbol{x}_{t=0}=\boldsymbol{x}_0 \,.
\end{equation}
Its solution defines the flow map $\boldsymbol{x}_t = \psi_t(\boldsymbol{x}_0)$.

The probability density evolves according to the instantaneous change-of-variables equation. The sample and its density
can therefore be propagated simultaneously by solving
\begin{equation}
  \od{}{t} \begin{bmatrix}
             \boldsymbol{x}_t \\
             \log q_t(\boldsymbol{x}_t)
           \end{bmatrix} = \begin{bmatrix}
                             \boldsymbol{v}_{t,\theta}(\boldsymbol{x}_t) \\
                             - \nabla \cdot \boldsymbol{v}_{t,\theta}(\boldsymbol{x}_t)
                           \end{bmatrix}
\end{equation}
with initial condition
\begin{equation}
  \begin{bmatrix}
    \boldsymbol{x}_t \\
    \log q_t(\boldsymbol{x}_t)
  \end{bmatrix}_{t=0} = \begin{bmatrix}
                          \boldsymbol{x}_0 \\
                          \log q_0(\boldsymbol{x}_0)
                        \end{bmatrix} \,.
\end{equation}
Forward integration generates samples and their densities, whereas the density of a prescribed endpoint can be evaluated
by integrating the same equations backwards. In practice, both operations are performed in approximated manner using a numerical ODE solver.

To obtain variables on the \Chili domain, the endpoint $\boldsymbol{x}_1$ is mapped component-wise from $\mathbb{R}^d$ to the open
unit hypercube $(0,1)^d$ using a sigmoid transform $\Sigma_d$. Let $G_0 : (0,1)^d \to \mathbb{R}^d$ denote a transformation
that maps uniform inputs to the base density $q_0$. The remapping introduced in Eq.~\eqref{eq:remap} is then
\begin{equation}\label{eq:remapping_chain}
  R_{\theta,N} = \Sigma_{d_N} \circ \psi_1 \circ G_0,\qquad d=d_N \,.
\end{equation}
The Jacobian of the sigmoid transformation is included when evaluating the resulting density $q_N$ on the unit
hypercube.

While maximum-likelihood training of a CNF requires ODE integration to evaluate the model density, CFM 
avoids ODE solves during training by directly regressing the vector field~\cite{Lipman:2023,Tong:2024}.
Let $p$ denote the target density in flow space. Using the independent coupling
\begin{equation}\label{eq:independent_coupling}
  \boldsymbol{x}_0 \sim q_0,\qquad \boldsymbol{x}_1 \sim p
\end{equation}
we sample $t \sim \mathcal{U}(0,1)$ and define the conditional straight-line interpolation
\begin{equation}
  \boldsymbol{x}_t = t \boldsymbol{x}_1 + (1-t) \boldsymbol{x}_0 \,.
  \label{eq:xt}
\end{equation}
Its conditional velocity is $\boldsymbol{x}_1-\boldsymbol{x}_0$, which leads to the regression objective
\begin{equation}\label{eq:cfm_objective}
  \mathcal{L}_{\text{CFM}}(\theta) = \mathbb{E}_{\substack{\boldsymbol{x}_0 \sim q_0,\;\boldsymbol{x}_1 \sim p\\
                                                           t \sim \mathcal{U}(0,1)}}
                                     \lVert \boldsymbol{v}_{t,\theta}(\boldsymbol{x}_t) - (\boldsymbol{x}_1-\boldsymbol{x}_0) \rVert_2^2\,.
\end{equation}
The minimiser recovers the marginal vector field associated with this interpolation and therefore defines a transport from $q_0$ to $p$.

In practice we perturb the interpolation point by Gaussian noise,
\begin{align}
  \boldsymbol{x}_t &= t \boldsymbol{x}_1 + (1-t) \boldsymbol{x}_0 + \boldsymbol{\varepsilon}\,,\\
  \boldsymbol{\varepsilon} &\sim \mathcal{N}(\mathbf{0}, \sigma_{\text{noise}}^2 \mathbf{1})\,,
\end{align}
which regularises the regression target in regions where the interpolation paths are densely packed. The conditional
velocity target $\boldsymbol{x}_1 - \boldsymbol{x}_0$ is unchanged.

The coupling in Eq.~\eqref{eq:independent_coupling} is independent: $\boldsymbol{x}_0$ and $\boldsymbol{x}_1$ are drawn from their respective
marginals with no correspondence between them. If instead a set of matched pairs $(\boldsymbol{x}_0, \boldsymbol{x}_1)$ is available, e.g.\
endpoints of trajectories generated by a previously trained flow, the same objective can be used with that coupling in
place of the independent one. The regression target $\boldsymbol{x}_1 - \boldsymbol{x}_0$ then reproduces the transport that generated the pairs,
rather than an arbitrary assignment between the two densities~\cite{Liu:2022,Tong:2024}.

In our application, direct samples from $p$ are unavailable. Instead, we draw $\boldsymbol{u}_N$ from a proposal density $q_N(\boldsymbol{u}_N)$,
transform the points to flow space, and multiply each loss term by the importance weight
\begin{equation}
  w(\boldsymbol{u}_N) = \frac{\mathcal{F}_N(\boldsymbol{u}_N)}{q_N(\boldsymbol{u}_N)}\,.
\end{equation}
Up to the common normalisation of the target density, this weighted expectation is equivalent to
Eq.~\eqref{eq:cfm_objective} with $\boldsymbol{x}_1 \sim p$. The weight is the event weight introduced in Eq.~\eqref{eq:event_weight}.
\section{Knowledge Transfer via Phase-Space Augmentation}
\label{sec:knowledge-transfer}
Evaluating high-multiplicity matrix elements is computationally
expensive, making the generation of training data for (ML-based) phase-space generators costly. To address this issue, we
propose a new training strategy that requires fewer matrix-element evaluations and yields models whose performance can surpass
that of models trained on larger, computationally more expensive datasets. 
The idea is to transfer knowledge which has already been learned by training on data for a lower-dimensional case to a
higher-dimensional case. Thereby the convergence of the higher-dimensional model can be accelerated, avoiding the
generation of costly training data.

In our application, the coordinate nesting of the \Chili parameterisation,
Eq.~\eqref{eq:coord_nesting}, allows us to capture essential parts of the $(N+1)$-particle distribution by simply
augmenting $N$-particle coordinate vectors with three additional coordinates. 
This is the one structural requirement the method places on the sampler. The transfer is meaningful only for a
phase-space parameterisation that follows the factorisation of Eq.~\eqref{eq:phi_n_plus_one_phi_one} and, more
specifically, one in which the inherited coordinates retain both their meaning and their ranges at the higher
multiplicity. Only then does a copied coordinate still describe the same kinematic degree of freedom, and only then is a
copied value guaranteed to lie in the admissible domain. The \Chili parameterisation satisfies this: the same particle
acts as the recoiler at every multiplicity, the gluons are generated in index order, and their coordinate ranges are set
by the cuts and the collider energy rather than by the multiplicity, cf.\ Sec.~\ref{subsec:PS-parametrization}. Beyond
this requirement the method is agnostic to the sampler being trained. It applies to Normalising Flows as well as to
classical adaptive algorithms such as \Vegas, although the specific implementation of the augmentation will differ.

Below, we describe how the lower-multiplicity data are augmented in this study. As a complication, the sampling
distributions are conditioned on the helicity configurations of the external particles and the number of non-vanishing
helicity states differs between multiplicities. Therefore, it is not enough to add new coordinate dimensions but one also
needs to augment the helicities. Otherwise, a large part of the target distribution would remain uncovered. It is important to
establish the augmented helicity labels before the added kinematic dimensions are filled. We show how this can be
achieved. Finally, we explain how the augmented data can be used to initialise a CNF for the next-higher multiplicity.

\subsection{Phase-Space Augmentation}\label{subsec:augmentation}
Assuming that a sampler for the $N$-particle phase space $\mathrm{\Phi}_{N}$ has been well-trained, a set of phase-space
points with distribution close to the target distribution can easily be generated. Note that at this point no weights
need to be generated and no matrix elements need to be evaluated, so the generation is cheap compared to evaluating the
integrand, and free if a suitable $\Phi_N$ dataset is already at hand.

Since the \Chili map $S_N$ acts component-wise and monotonically on the hypercube variables, cf.\
Eqs.~\eqref{eq:y_map}--\eqref{eq:pT_dist_top}, copying, shuffling and re-sampling individual coordinates commute with it. We
therefore present the augmentation in terms of the \Chili coordinates $\boldsymbol{z}_N$, which carry direct kinematic
meaning, while the actual implementation operates equivalently on the unit-hypercube variables $\boldsymbol{u}_N$.

Based on the \Chili coordinate nesting, the set of
$d_{N}$ $N$-particle coordinates $\boldsymbol{z}_N$ of an event can be copied from $\mathrm{\Phi}_{N}$ to the corresponding 
dimensions in the $\mathrm{\Phi}_{N+1}$ without losing their meaning:
\begin{equation}
    \mathbf{z}_N =
    \overset{\raisebox{8pt}{$\textstyle \mathrm{\Phi}_{N}$}}%
             {\usebox{\firstvecbox}}
    \;\xrightarrow{\text{copy}}\;
    \overset{\raisebox{8pt}{$\textstyle \mathrm{\Phi}_{N+1}$}}%
             {\topalignedbmatrix{
                z_N^{(1)} \\
                z_N^{(2)} \\
                \vdots \\
                z_N^{(d_{N})} \\
                \ast \\
                \vdots \\
                \ast
             }}
    \quad .
\end{equation}
The three additional dimensions, corresponding to the new resolved gluon, are denoted by an asterisk above. From here on we
collectively refer to them as $\tilde{\boldsymbol{\xi}} = (\tilde{p}_{n+1,\perp}^2, \tilde{y}_{n+1}, \tilde{\varphi}_{n+1})$. 
Filling these additional dimensions creates an augmented vector
$\tilde{\boldsymbol{z}}_{N+1} = (\boldsymbol{z}_N, \tilde{\boldsymbol{\xi}})$, or equivalently $\tilde{\boldsymbol{u}}_{N+1} =
S_{N+1}^{-1}(\tilde{\boldsymbol{z}}_{N+1})$ on the unit hypercube, which represents a candidate
$(N+1)$-particle phase-space point that may fail the cuts or fall outside the physical domain. 
The distribution of the augmented vectors is
expected to capture essential parts of the structure of the actual $(N+1)$-particle distribution:
\begin{equation}
  q_{\text{aug}} \approx p_{N+1}\,,
\end{equation}
although the prescription cannot reproduce the distribution exactly. 
To fill the appended coordinates, we propose two alternative approaches.

\begin{figure*}[tbp]
    \centering
    \begin{subfigure}[t]{0.47\textwidth}
        \centering
        \includegraphics[width=.7\linewidth]{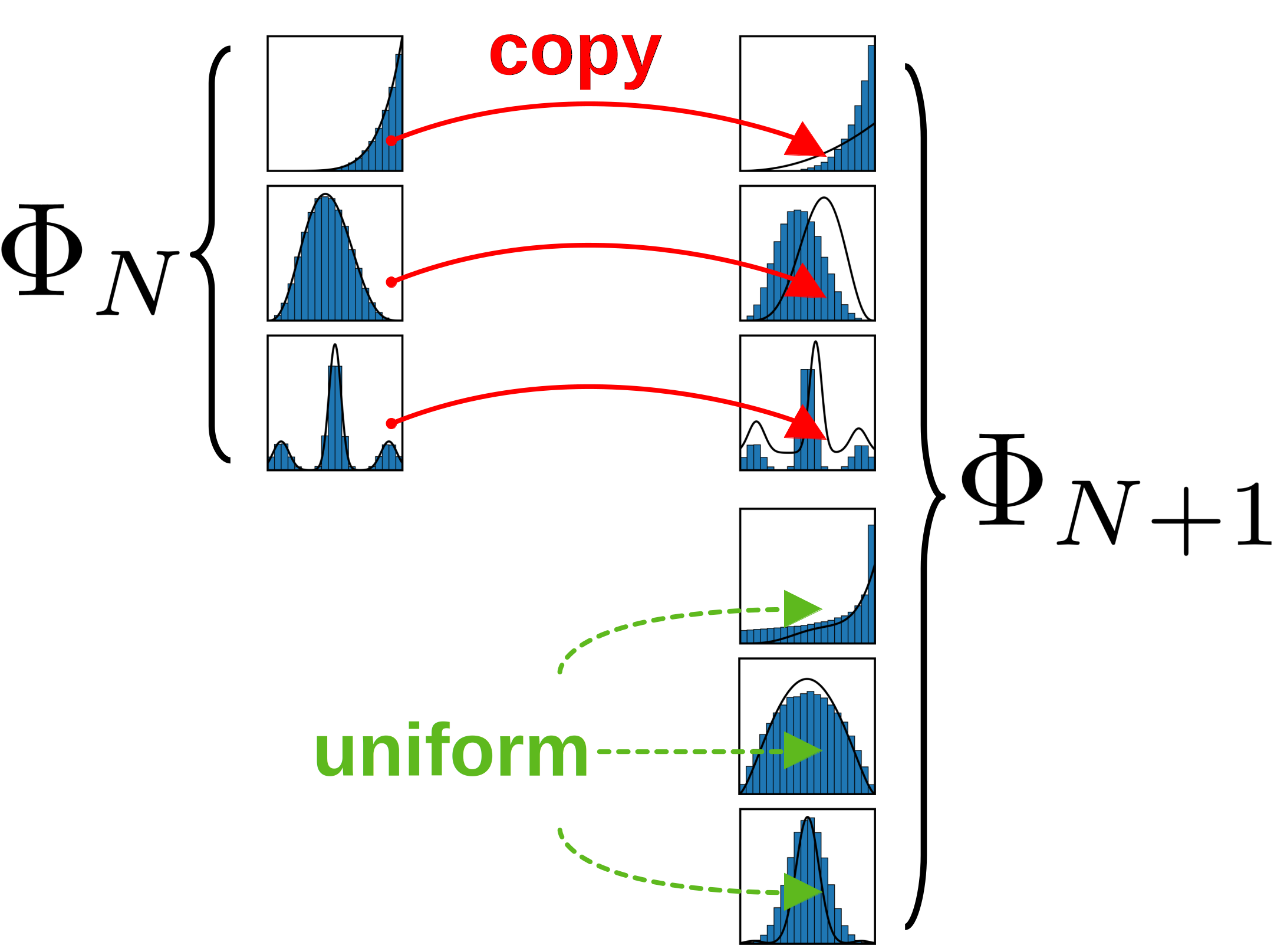}
        \caption{Uninformed augmentation: the missing dimensions are
                 filled with uniform noise.\\}
        \label{fig:aug-uninformed}
    \end{subfigure}
    \hfill
    \begin{subfigure}[t]{0.47\textwidth}
        \centering
        \includegraphics[width=.7\linewidth]{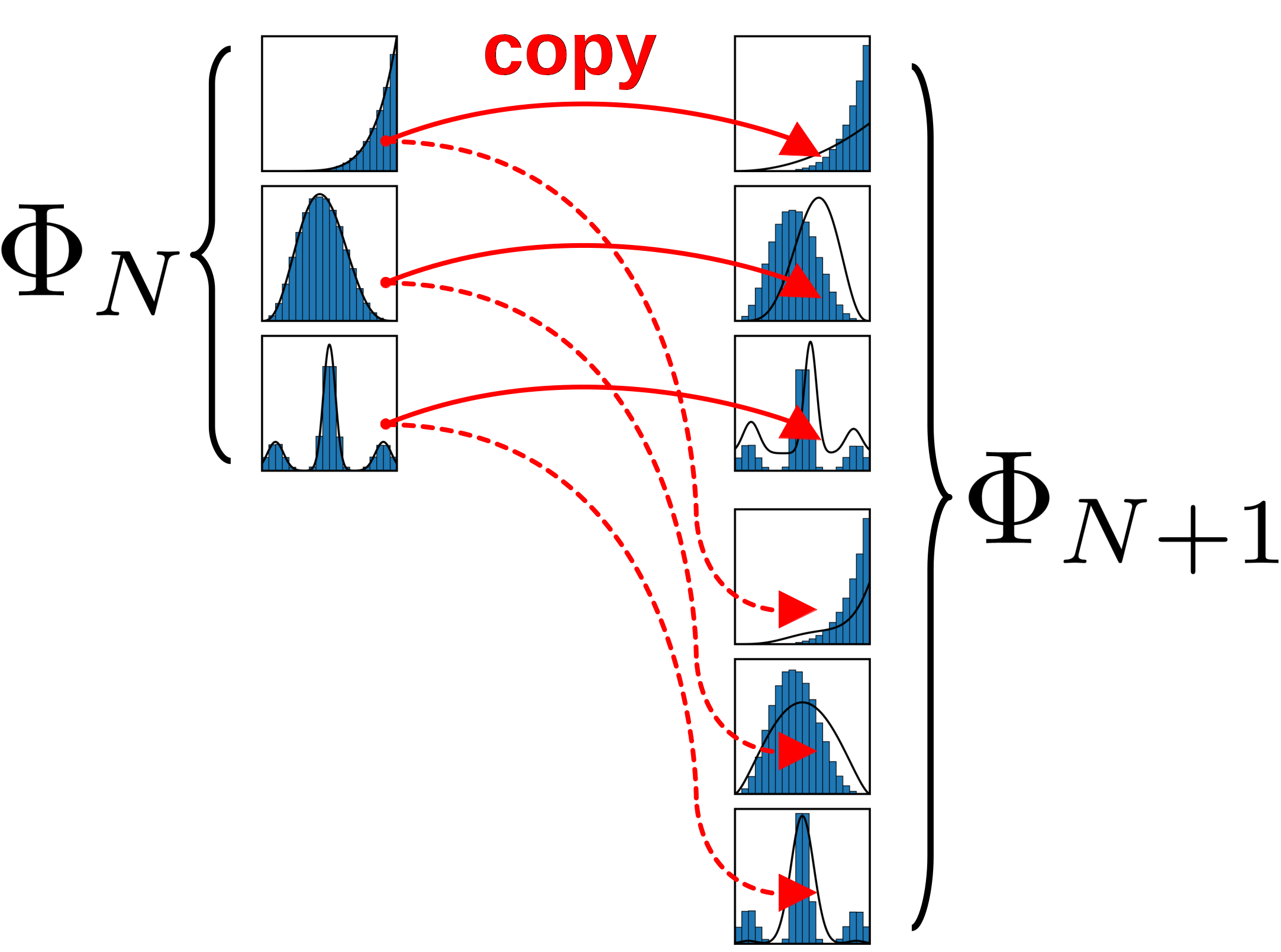}
        \caption{Informed augmentation: the extra dimensions are filled
                 with distributions that reflect the expected ones.}
        \label{fig:aug-informed}
    \end{subfigure}
    \caption{Uninformed (\ref{fig:aug-uninformed}) and informed (\ref{fig:aug-informed}) ways to augment an
      $N$-particle to an $(N+1)$-particle phase space. In both cases, the phase-space mapping is 
      applied to the inserted coordinates.}
    \label{fig:data-augmentation-schemes}
\end{figure*}

In the simplest case, no knowledge about the new dimensions is assumed and the respective components of
$\tilde{\boldsymbol{u}}_{N+1}$ are drawn uniformly and mapped to the \Chili variables, as shown in
Fig.~\ref{fig:aug-uninformed}. We refer to this procedure as \emph{uninformed augmentation}. 
The uninformed approach makes no assumptions about the distributions in the additional phase-space 
dimensions, and is therefore the most generic form of the method.

When expert knowledge about the remaining dimensions is available, an \emph{informed augmentation} can be used, cf.\
Fig.~\ref{fig:aug-informed}. The additional dimensions can be filled with distributions that roughly match
the expected distributions. In our case, the extra dimensions correspond to degrees of freedom of an extra particle,
parameterised the same way as the other particles. As a simple example, we choose to copy the coordinates of the $n$-th
gluon in $\boldsymbol{z}_N$ to the $(n+1)$-th position in $\tilde{\boldsymbol{z}}_{N+1}$. To avoid spurious correlations
between the two gluons, the coordinates are decorrelated by shuffling: each of the three new coordinates is chosen
randomly from all available events without replacement. This preserves only their marginal distributions. The first $d_N$ coordinates in $\tilde{\boldsymbol{z}}_{N+1}$, directly copied from
$\boldsymbol{z}_N$, are not shuffled in order to keep their correlations. This yields a deliberately simple procedure;
more sophisticated informed augmentations are certainly conceivable.

Note that in neither of the two augmentation schemes do we filter the augmented
$\mathrm{\Phi}_{N+1}$ points for their kinematic validity or phase-space cuts.
All augmented points are kept for training the initial flow map.

Since a valid $N$-particle point need not remain valid once a gluon is added, cf.\ Section~\ref{subsec:PS-parametrization}, we regard the
lower-multiplicity distribution as providing approximate, rather than complete, coverage of the projected
higher-multiplicity one. Regions not covered by this initial transfer can be learned during the subsequent
importance-sampling iterations.

Furthermore, since the three additional dimensions are simply added without affecting the $d_N$ dimensions 
before, the proposed phase-space point $\tilde{\boldsymbol{z}}_{N+1}$ does not respect the fixed global event 
kinematics. Most notably, this changes the total overall momentum. 

\subsection{Conditional Augmentation}\label{subsec:conditional_augmentation}
Since \Pepper uses helicity sampling, the optimal sampling distribution depends on the helicity configuration of the 
external particles. Helicity combinations for which the matrix-element evaluation
yields zero are excluded from all datasets. By default, \Pepper does not take advantage of the conditioning
and simply uses one \Chili mapping in combination with a single \Vegas remapping for all helicity configurations of a
given partonic process. With a Normalising Flow remapping, however, it is easy to condition the sampling distribution on
the helicity configurations and learn the correlations between the phase-space variables and the helicity configuration, as shown
in Ref.~\cite{Bothmann:2025lwg}. In the context of augmentation, one has to be careful, though, since the addition of a
particle increases the number of available helicity states. For the processes considered here, when an $N$-particle
dataset is augmented to an $(N+1)$-particle
dataset, the number of non-vanishing helicity
states always doubles. So the data needs to be augmented to twice the number of helicity states, besides adding three kinematic
dimensions. Fortunately, this turns out to be straightforward in \Pepper.

\Pepper enumerates all $2^{N+2}$ configurations, disables those whose colour-summed amplitude vanishes, and
labels the remainder consecutively. We use this compacted label $c \in {0, \dots, n_{\text{hel}, N} - 1}$ as the
conditioning variable. For $gg \to t \bar{t} + ng$ no configuration vanishes and $n_{\text{hel}, N} = 2^{N+2}$, whereas
for $d \bar{d} \to e^+ e^- + ng$ helicity conservation along the quark and lepton lines leaves $n_{\text{hel}, N} =
2^{N}$ of $2^{N+2}$. 

The additional gluon is appended at the end of the particle list, so its helicity is the most significant bit of
\Pepper's internal label, both of its values are allowed, and its presence does not change which of the remaining
configurations vanish. The enabled set at multiplicity $(n+1)$ is therefore the enabled set at $n$, together with its image
under setting the new bit, and since the consecutive labelling preserves the internal ordering, each event
$[\boldsymbol{z}_N,c]$ is duplicated with the label unchanged in one copy and shifted by $n_{\text{hel}, N}$ in the other:
\begin{equation}\label{eq:helicity_doubling}
  \bigl[\boldsymbol{z}_N, c\bigr]
  \;\longrightarrow\;
  \begin{cases}
    \bigl[(\boldsymbol{z}_N, \tilde{\boldsymbol{\xi}}),\; c\bigr]\,, \\[3pt]
    \bigl[(\boldsymbol{z}_N, \tilde{\boldsymbol{\xi}}),\; c + n_{\mathrm{hel},N}\bigr]\,,
  \end{cases}
\end{equation}
where $\tilde{\boldsymbol{\xi}}$ is drawn according to Sec.~\ref{subsec:augmentation}, and
the shuffle underlying its construction is performed within the same label class.
This doubles the dataset and covers all $n_{\text{hel}, N+1} = 2 n_{\text{hel}, N}$ configurations of the $(N+1)$-particle system.

This scheme assumes that the kinematic behaviour
of the original $N$ particles, given some helicity assignment, remains essentially 
unchanged when an extra gluon is added, as long as the coordinates $\boldsymbol{z}_N$ are kept fixed. Nevertheless, it
approximates the marginal distributions of the $d_{N}$
dimensional projection of the $\mathrm{\Phi}_{N+1}$ phase space adequately.

In \Pepper the helicity configuration is not stored explicitly but selected from a uniform random number against the
cumulative selection weights $\alpha_c$. The augmented labels of Eq.~\eqref{eq:helicity_doubling} are therefore
re-encoded against the $(N+1)$-particle weights before the dataset is passed back to \Pepper. Since no $(N+1)$-particle
statistics are available at this point, we initialise these by splitting each inherited weight between the two labels it
gives rise to, $\alpha_c^{(N+1)} = \alpha_{c+n_{\text{hel},N}}^{(N+1)} = \frac{1}{2} \alpha_c^{(N)}$, which preserves
both the normalisation and the relative importance learned at multiplicity $N$. From the first iteration onwards the
weights are updated as described above.

\subsection{Training on Augmented Data}\label{subsec:training_augmentation}
Augmented data can be used to initialise a model for the $(N+1)$-particle process by generating matched pairs
$(\boldsymbol{x}_0, \boldsymbol{x}_1)$ and regressing against them as described in Sec.~\ref{subsec:ml-background}. Both
$\boldsymbol{x}_0$ and $\boldsymbol{x}_1$ are constructed by augmenting lower-dimensional data. 
The flow of Sec.~\ref{subsec:ml-background} operates neither on the \Chili coordinates nor on the unit hypercube, but on
$\mathbb{R}^{d_{N+1}}$, cf.\ Eq.~\eqref{eq:remapping_chain}. The augmented points of Sec.~\ref{subsec:augmentation} are therefore pushed forward accordingly,
\begin{align}
  \boldsymbol{x}_1 &= \Sigma_{d_{N+1}}^{-1}\big(\tilde{\boldsymbol{u}}_{N+1}\big),\\
  \tilde{\boldsymbol{u}}_{N+1} &= S_{N+1}^{-1}\big(\tilde{\boldsymbol{z}}_{N+1}\big)\,,
\end{align}
and the $\boldsymbol{x}_0$ are matched base points constructed as follows.
%%%
For both the uninformed and the informed augmentation, the $d_N$ components inherited from $\Phi_N$ are copied at $t=0$
and at $t=1$ alike, so that their pairing stays intact. This implies that the points drawn from the base distribution at
$t=0$ need to be saved, which would not be necessary for standard CFM training. The treatment of the three
added components differs between the two approaches. In the uninformed case, the new components of $\boldsymbol{x}_1$
are drawn uniformly and the corresponding components of $\boldsymbol{x}_0$ independently from the ODE base
distribution, so only the independent coupling between them is learned. In the informed case, both are copied from a
previously generated gluon, so that the coupling that produced that gluon's trajectory is inherited directly. 

The goal
here is to learn the coupling as a sensible initialisation for the model. Notably, the goal is not yet to find the
optimal map for sampling the target distribution and for that reason no target-function evaluations are necessary at
this stage. Thus, the training on augmented data is comparatively cheap while resulting in better initialisations than
training from scratch on uniformly sampled data with proper weights. This is due to the reuse of previously learned information.
Since no weights are calculated, this benefit comes with a significant reduction in computational costs. This effect
becomes larger with increasing multiplicity due to the scaling of the evaluation costs of the matrix elements.

After initialising the model with the augmented data, the model can subsequently be trained in the standard way. For
CNFs, the training procedure with CFM is discussed in Sec.~\ref{subsec:ml-background}. We describe the
practical implementation for our setting in Sec.~\ref{sec:nw_architecture}.
\section{Application to Multijet-Production Processes}
\label{sec:application_to_multijet_production}
\begin{figure}[tbp]
    \centering
    \includegraphics[width=\linewidth]{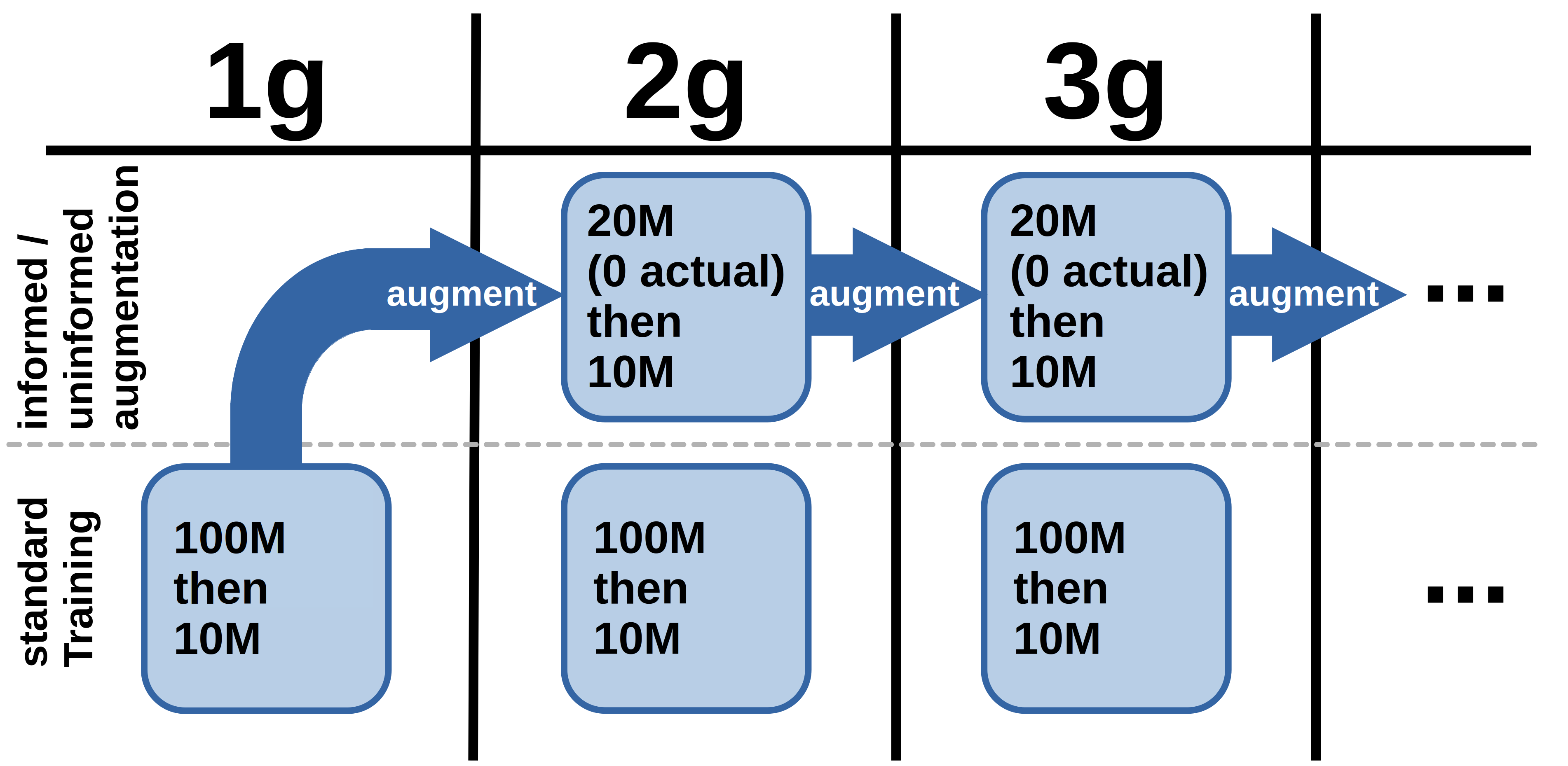}
    \caption{Build up of the training stack and initial $n g$ dataset sizes.}
    \label{fig:training-stack}
\end{figure}

In this section we describe the application of our data-augmentation approach in the iterative 
training of CNFs for phase-space point generation for multijet-production processes. To this end,
we consider $d\bar{d} \to e^+e^- + n g$ and $gg \to t\bar{t}+n g$ production
with $n\leq 5$ in proton--proton collisions at $\sqrt{s} = 13\,\text{TeV}$.
Our training proceeds sequentially: after a CNF for a given gluon-multiplicity process is trained, 
we augment a generated dataset of phase-space points to the next higher multiplicity, and train the next
CNF.  We refer to this training procedure as \emph{stacked training}, see Fig.~\ref{fig:training-stack}.

\subsection{Calculational Tools}
\label{subsec:calculational_tools}
In this application, the GPU compatible event-generation framework \Pepper~\cite{Bothmann:2023gew}, 
alongside its colour-summed matrix element implementation and \Chili~\cite{Bothmann:2023siu} phase-space mapping is used. 
For a detailed description of the phase-space parameterisation see Sec.~\ref{subsec:PS-parametrization}. Helicity states 
are sampled in proportion to their relative contribution to the variance of the total cross section. For the parton distribution 
functions appearing in the hadronic cross sections, cf.\ Eq.~\eqref{eq:hadroic-cross-section},
we employ the NNPDF3.0~\cite{NNPDF:2014otw} set, accessed through the LHAPDF6 library~\cite{Buckley:2014ana},
that furthermore provides the running QCD coupling. The factorisation and renormalisation scales are set dynamically
based on the event kinematics. For lepton-pair production, they are set to
\begin{align}
    \mu_{\text{F}}^2 = \mu_{\text{R}}^2 = \frac{{H}_{\text{T}}^{'2}}{2} &= \frac12 \left( m_{\perp, \ell\ell} + \sum_{\text{gluons}} p_{\perp, g} \right)^2  \\
    \text{with}\quad m_{\perp, \ell\ell} &= \sqrt{m^2_{\ell\ell} + p^2_{\perp, \ell\ell}}\,, 
\end{align} 
whereas for top-pair production, they are set to 
\begin{equation}
  \mu_{\text{F}}^2 = \mu_{\text{R}}^2 = \frac{H_{\text{T}}^2}{2} = \frac12\left( p_{\perp, t} + p_{\perp, \bar{t}} + \sum_{\text{gluons}} p_{\perp, g} \right)^2.
\end{equation}

The electroweak mixing angle and coupling constant are set to $\sin^2\Theta_{\text{w}} = 0.23155$,
and $\alpha_{\text{em}} = 128.802223295^{-1}$. The $Z$-boson mass and its width are set to $m_{Z} = 91.1876\,\text{GeV}$,
and $\Gamma_Z = 2.4952\,\text{GeV}$, respectively. Furthermore, the top-quark mass is set to $m_{t} = 172.5\,\text{GeV}$.

On the final-state \textit{jets}, i.e.\ gluons, we apply a generic set of cuts:
\begin{align}
  p_{\perp,j} &\geq 30\,\text{GeV}\,,\label{eq:jetcut_pt}\\
  |y_j| &\leq 5\,,\label{eq:jetcut_eta}\\
  \Delta R_{jj} &=  \sqrt{(\Delta y_{jj})^2 + (\Delta\varphi_{jj})^2} \geq 0.4\,,\label{eq:jetcut_dR}
\end{align}
with $\Delta y_{jj}$ and $\Delta\varphi_{jj}$ the rapidity and azimuthal separation of any pair of jets, respectively.
Consistently applying cuts on the kinematics of all (QCD) particles helps to avoid support mismatches
between the set of
augmented points and the full target distribution in $\mathrm{\Phi}_{N+1}$. For the lepton-pair production processes,
we furthermore use universal cuts on the non-QCD particles, independent of the gluon multiplicity:
\begin{align}
    66\,\text{GeV}&\leq m_{\ell\ell} \leq 116\,\text{GeV}\,.\label{eq:lepton_system_cut_mass}
\end{align}

\subsection{Network Architecture and Training Parameters}\label{sec:nw_architecture}
The CNF models are implemented in \pytorch~\cite{2019arXiv191201703P} and ODE integration is done using
\torchdyn~\cite{2020arXiv200909346P}. The implemented model is a simple 
multilayer perceptron with four hidden layers, and a phase-space dimension 
dependent input and output layer. The CNF is trained using the CFM 
objective, cf. Eq.~\eqref{eq:cfm_objective}, where each event's conditional
vector-field prediction is weighted by the corresponding event weight. 

For each training step the model receives three inputs: the interpolated point
$x_t$, the interpolation time $t$, and an embedding of the helicity condition 
of the target event. The total number of trainable 
parameters, including the helicity weights, is of order one million, see 
Tab.~\ref{tab:trainable-params}. A full list of architectural choices is given 
in Tab.~\ref{tab:model-building-blocks}. 

For model-parameter optimisation we use 
\textsc{AdamW}~\cite{2017arXiv171105101L} with a constant learning rate of $10^{-3}$. 
To improve training stability, \textsc{AMSgrad}~\cite{j.2018on} is used in \textsc{AdamW}. 
At the start of every new iteration the optimiser is re-initialised because 
statistical properties of the newly generated datasets, such as the event weights, 
change, as the model improves and the event-weight distribution becomes 
progressively narrower across iterations. The helicity weights used to generate a 
dataset are updated after each iteration by the normalised sum of MC weights of all 
events that belong to each helicity class.

Reference models are trained as in Ref.~\cite{Bothmann:2025lwg}, referred to 
as \textit{standard training} in what follows.  Accordingly, for each process and
every gluon multiplicity, the reference models are trained for 600 epochs on 
a dataset of roughly $100\text{M}$ uniformly sampled weighted events.
After the initial training phase, 
the current models generate a new dataset of about $10\text{M}$ weighted events, 
which is then used for a second 600-epoch training phase. In the dataset generation, 
helicity conditions are generated according to their corresponding 
weights. The ODE integration in \textsc{TorchDyn} is accomplished with the DormandPrince45
solver~\cite{DORMAND198019}, with absolute and relative tolerances set to $10^{-4}$. The 
model-generated points and their corresponding model weights are written to disk, then read 
in by \textsc{Pepper}. \textsc{Pepper} evaluates the PDFs, phase-space mapping, and matrix element 
at each input point, computes the corresponding MC weight, cf. Eq.~\eqref{eq:event_weight}, and 
stores the result in the LHEH5 event format~\cite{Hoche:2019flt,PhysRevD.109.014013}. 
This iterative training cycle is 
repeated until either an early-stopping criterion is met or a maximum of 50 
iterations is reached. Training stops when the newly generated dataset has a relative  
effective sample size $\frac{N_{\text{eff}}}{N_{\text{ev}}}$, and unweighting efficiencies 
$\epsilon_{1\%}$ and $\epsilon_{0.1\%}$ that improve by less than 5\% relative 
to the previously generated dataset.
Throughout model training, a batch size of $2^{19}$ is used. For the reference models,
this training procedure is repeated for all gluon multiplicities, both, for lepton- and
top-pair production. Accordingly, the large initial dataset of $100\text{M}$ uniformly-sampled
events needs to be generated for all gluon multiplicities individually.

To evaluate the benefit of our stacked-training strategy, both for Drell--Yan and top-quark
production, we construct two training stacks, one using informed augmentation, the other using
uninformed augmentation. 

In either stacked training, we augment the reference $X+1g$-standard training's 
final $\sim 10\text{M}$-event dataset to the next higher 
multiplicity $X+2g$. Rather than generating fresh $\Phi_N$ points as described in Sec.~\ref{subsec:augmentation}, we
simply reuse this dataset: its points have already been paid for during the $X+1g$ training, and only their phase-space
coordinates and helicity labels enter the augmentation, so the associated weights are irrelevant here. These augmented
datasets serve as initial datasets for the iterative training of 
the respective $2g$ models, omitting the need to generate the large uniformly sampled 
initial datasets. A detailed description of the 
data augmentation can be found in Sec.~\ref{sec:knowledge-transfer}.

Since the augmented dataset only approximates the true $(N+1)$-particle distribution, we 
increase the Gaussian noise added to the interpolation point $x_t$ in the first-iteration 
training on the augmented dataset from the usual standard deviation of $\sigma_\text{noise}=10^{-4}$
to $\sigma_\text{noise}=10^{-1}$, assuming a mean of zero. For the following iterations we resort to 
$\sigma_\text{noise}=10^{-4}$ as used for the reference model.

\subsection{Results}
\label{subsec:results}
In this section, we present the reduction in computational cost 
achieved with the stack of augmented trainings against reference benchmarks.
We denote the informed augmentation by \emph{Stack cp.} and \emph{Stack unif.} refers to
the uninformed variant. To ensure a fair comparison, we show two benchmark trainings.
The \emph{Standard Training 100M}, for brevity denoted as ST100 in the text, follows
exactly the training procedure of Ref.~\cite{Bothmann:2025lwg}, summarised in Sec.~\ref{sec:nw_architecture},
including the large initial datasets of $\sim 100$M uniformly sampled events\footnote{The number of
non-zero events is determined by the uniform dataset's phase-space efficiency, which especially for
the higher gluon multiplicities, is significantly smaller than $100\%$.}. 
In a second benchmark, \emph{Standard Training 20M}, denoted as ST20 below, the 
initial uniform dataset is reduced to contain exactly the same number of non-zero events as 
the informed training stack. In the augmented trainings, as the size of the $(N+1)$-particle initial dataset 
is dictated by the phase-space efficiency of the augmented $N$-particle dataset, these are
slightly different for the informed and uninformed augmentation.

Both augmentation approaches show essentially the same reduction in training cost and a
similar final model performance. As in our application expert knowledge about the
additional phase space dimensions is available, we concentrate on the comparison of the
benchmark trainings to the informed augmentation. A comparison between the two augmentation
methods for the highest gluon-multiplicity channels can be found in Appendix~\ref{apx:additional_plots},
explicit results are quoted in Tab.~\ref{tab:performance_5g}.

\begin{figure*}[htpb]
    \centering
    \includegraphics[width=0.9\textwidth]{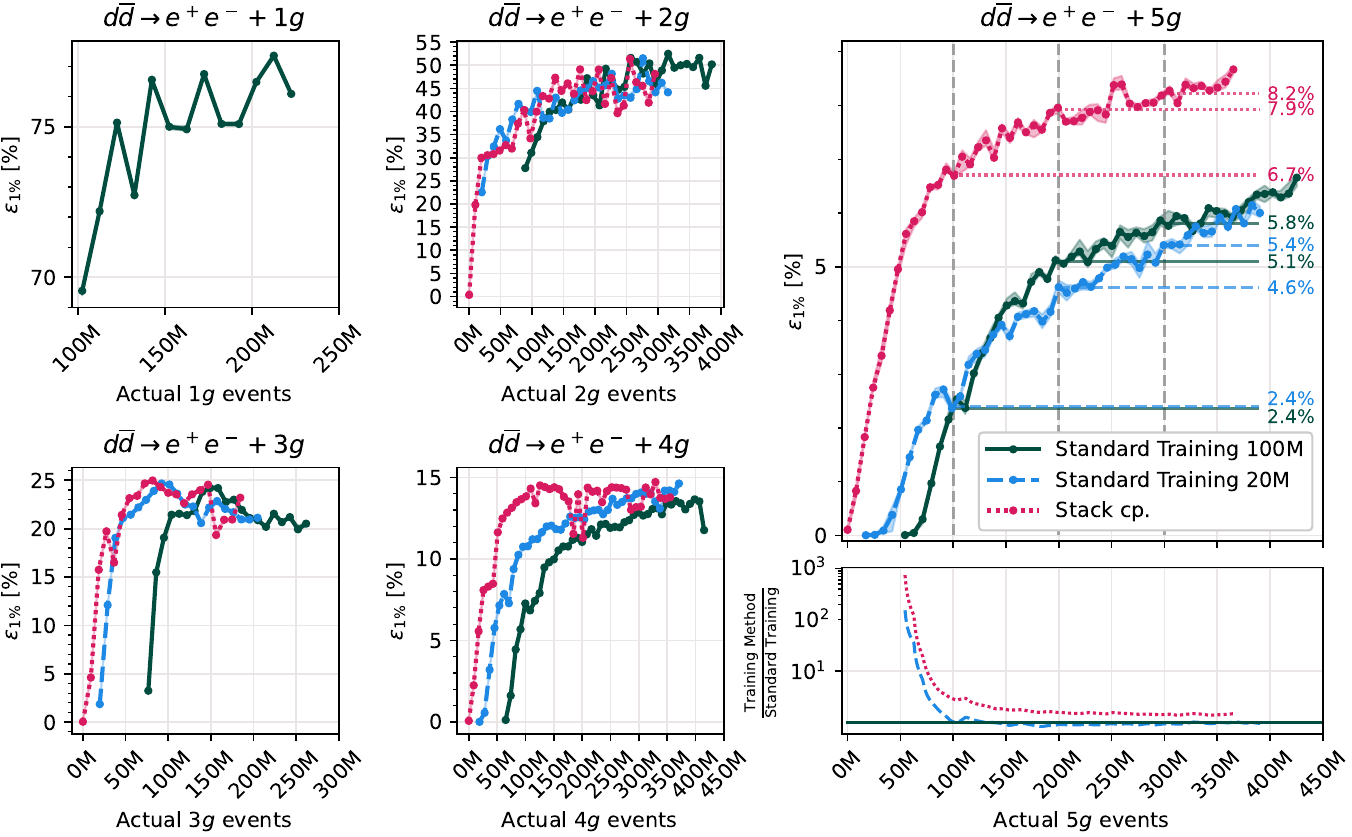}
    \caption{1\% overweight unweighting efficiencies for the 
    $d\bar{d} \to e^+e^- + n g$ stack, comparing the standard 
    training benchmark, standard training with only $\sim20$M initial 
    uniformly sampled points, and the informed-augmentation stack.}
    \label{fig:ddz_1_percent_uw_eff}
    \vspace*{5pt}
    \centering
    \includegraphics[width=0.9\textwidth]{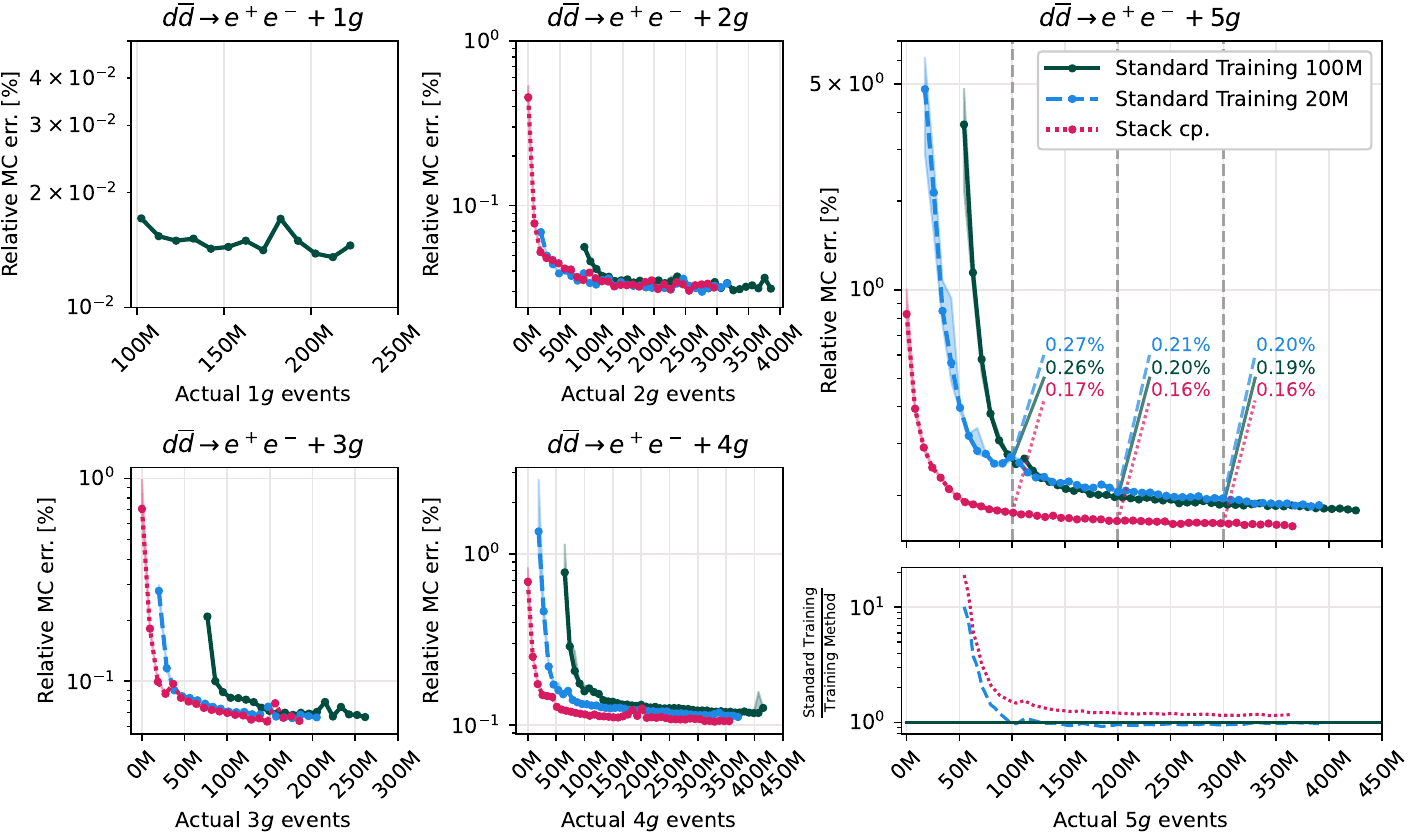}
    \caption{Relative MC integration error for the 
    $d\bar{d} \to e^+e^- + n g$ stack, comparing the standard 
    training benchmark, standard training with only $\sim20$M initial 
    uniformly sampled points, and the informed-augmentation stack.
    }
    \label{fig:ddz_mc_err}
\end{figure*}

\begin{figure*}[t]
    \centering
    \includegraphics[width=0.9\textwidth]{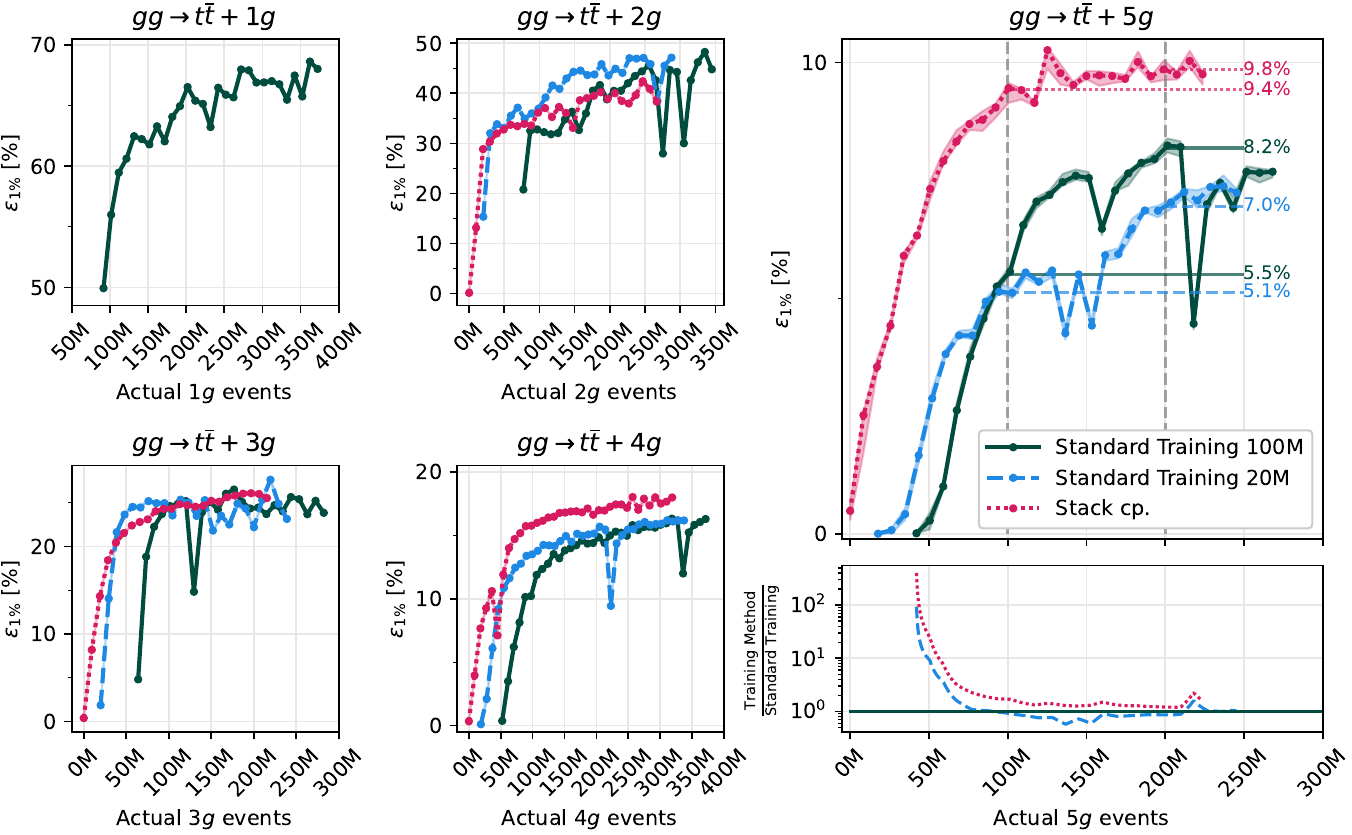}
    \caption{1\% overweight unweighting efficiencies for the whole 
    $gg \to t\bar{t} + n g$ stack, comparing the standard 
    training benchmark, standard training with only $\sim20$M initial 
    uniformly sampled points, and the informed-augmentation stack.}
    \label{fig:ggtt_1_percent_uw_eff}
    \vspace*{5pt}
    \centering
    \includegraphics[width=0.9\textwidth]{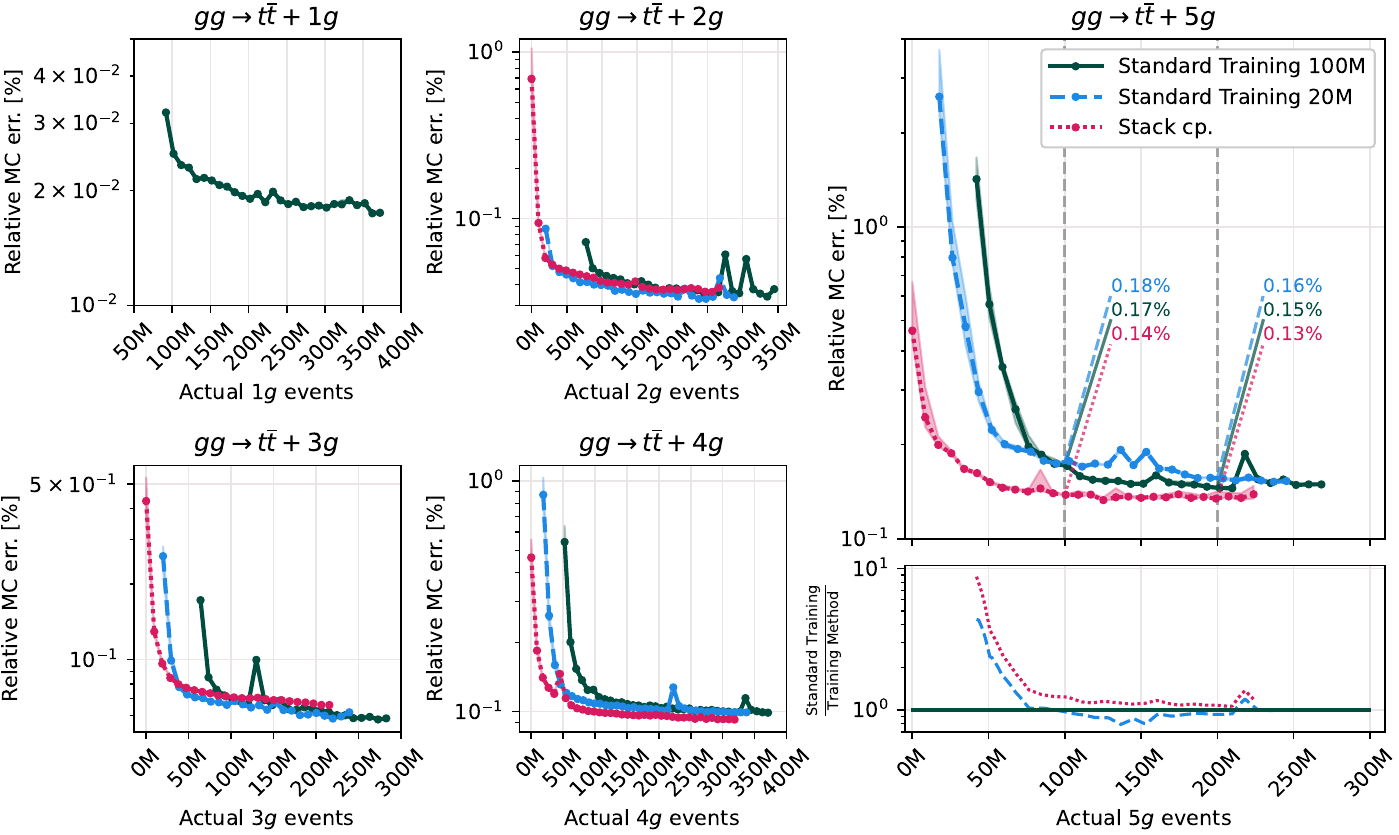}
    \caption{Relative MC integration error for the 
    $gg \to t\bar{t} + n g$ stack, comparing the standard 
    training benchmark, standard training with only $\sim20$M initial 
    uniformly sampled points, and the informed-augmentation stack.
    }
    \label{fig:ggtt_mc_err}
\end{figure*}

To assess the model performance, we consider multiple figures of merit, including the phase-space
efficiency, the error of the integral estimate, the effective sample size and the unweighting
efficiencies $\epsilon_{0.1\%}$ and $\epsilon_{1\%}$, cf. Tab.~\ref{tab:performance_5g}. For brevity,
we focus the discussion on $\epsilon_{1\%}$, presented for $d\bar{d}\to e^+e^-+ n g$ in Fig.~\ref{fig:ddz_1_percent_uw_eff} 
and $gg \to t\bar{t}+n g$ in Fig.~\ref{fig:ggtt_1_percent_uw_eff}, and the relative MC integration error
$\sigma_{\text{rel}} = \frac{\sigma_{\text{error}}}{\sigma}$ in Figs.~\ref{fig:ddz_mc_err} and \ref{fig:ggtt_mc_err}.
The performance metrics are shown in dependence of the total number of actual $n g$ events each model
has used during training. Accordingly, a training-cost reduction is achieved, when a model attains the same,
or superior, performance with respect to the benchmark models with fewer events.

All metrics are evaluated on ten independent datasets of 700k events. Each data point in the plots corresponds
to the median of the ten evaluations. The shaded band shows the middle 50\% of the data, from the 25th to 
the 75th percentile across the ten repetitions. 

For phase spaces with up to three gluons, the benchmark and the augmented trainings show similar 
performance when trained on a comparable amount of events. However, a clear advantage 
of the augmentation can be seen for final states with four or more gluons. Because each added gluon
in $X + n g$ processes contributes three additional dimensions, the fraction of phase-space dimensions 
that start with a well-initialised proposal distribution grows 
as $\frac{4 + 3 n}{7 + 3  n}$, approaching unity for large $n$. This accelerates the models'
learning and therefore yields higher final performance when trained on the same number of iterations
as the  benchmark trainings.
Additionally, since the models train on statistically richer datasets, they converge in fewer iterations, which further
reduces the cost of network training and of the ODE-based dataset generation. This saving is not reflected in the
matrix element cost metric used in the figures.

\paragraph{\boldmath$d\bar{d} \to e^+e^- + 4 g / 5 g$}
In Drell--Yan production with four gluons, the baseline trainings as well 
as augmentation-training converge to similar final performances in $\epsilon_{1\%}$, 
see Fig.~\ref{fig:ddz_1_percent_uw_eff}. However, to reach the ST100's peak unweighting
efficiency of $\epsilon_{1\%}=13.7\%$, 
the augmented training requires only 23\% of the actual four-gluon data, while ST20 needs about 64\%. 
The same efficiency gain is reflected in $\sigma_{\text{rel}}$, cf. Fig.~\ref{fig:ddz_mc_err},
where the augmentation training attains the smallest error achieved by the standard training
with only a fraction of actual four-gluon training events. 

Next, we compare the augmented training to the benchmarks in the case of Drell--Yan plus five 
gluons. Here, the phase space is 19 dimensional, making it, alongside the corresponding top-pair
production phase space, the highest-dimensional learning problem in this work. Accordingly, we expect
the gain in training efficiency to be largest. To this end, we present ratio plots comparing the
performance of the informed stack, and ST20 to ST100 benchmark. 
For $\epsilon_{1\%}$, cf. Fig.~\ref{fig:ddz_1_percent_uw_eff}, the informed training saturates faster
and at a higher value than the benchmark trainings. The ST100 training starts at around 50M events,
given a phase-space efficiency of about 50\% only. At this point, $\epsilon_{1\%}$ of the augmented
training is almost three orders of magnitude larger than the ST100's. The informed augmentation
reaches the maximal $\epsilon_{1\%}$ of ST100, while training on 22\% of actual five-gluon data only.
This is also reflected in $\sigma_{\text{rel}}$. Here, only about 20\% of training data needs to be
generated in the augmentation-training to reach the final performance of the standard training,
see Fig.~\ref{fig:ddz_mc_err}.  In conclusion, the augmented training cuts down the computational
cost of matrix-element evaluation by almost 80\%.

To test whether the increase in training efficiency is merely a result of using a 
smaller-sized initial dataset, therefore training on self-sampled data earlier, we compare the two
standard trainings. Here, both $\epsilon_{1\%}$ and  $\sigma_{\text{rel}}$ show that, ST20 is outperformed
by ST100 past $\sim150$M training events, cf. Fig.~\ref{fig:ddz_1_percent_uw_eff} and
Fig.~\ref{fig:ddz_mc_err}, respectively. This demonstrates that switching to self-sampled data earlier
in training does not provide a significant acceleration of convergence beyond the very start of the
training. In contrast, starting from a statistically richer initial dataset, as the informed augmentation
does, yields a clear benefit. 

Next, we discuss the final model performance of the different training methods, collated in
Tab.~\ref{tab:performance_5g}. The final models obtained with the augmentation strategies consistently
outperform the benchmark trainings in all considered metrics, except for the phase-space efficiency. 
Here, ST20 is marginally better, however, by less than 1\% relatively. In $\epsilon_{1\%}$, the
informed-augmentation training outperforms ST100 by 28\% relatively. % 27.99
Improvements are even larger when compared to the training strategy with the 
same amount of initial data, i.e. ST20. Here, $\epsilon_{1\%}$ is improved by 42\% relatively. % 42.03 
Furthermore, $\sigma_{\text{rel}}$ is 16\% smaller in the informed-augmentation training than in % 16.16
the two benchmarks. 

\paragraph{\boldmath$gg \to t\bar{t} + 4 g / 5 g$}
We now turn the discussion to top-pair production, starting out with the
four-gluon final state. For $\epsilon_{1\%}$ the informed-augmentation training
reaches the ST100's peak performance while training on 39\% of the actual four-gluon
events only, see Fig.~\ref{fig:ggtt_1_percent_uw_eff}. Furthermore, the final $\epsilon_{1\%}$
is improved by 11\% relatively. When comparing the \emph{uninformed}-augmentation training to ST100,
the final $\epsilon_{1\%}$ is improved by 16\% relatively, see Tab.~\ref{tab:performance_5g}.
Also for the $\sigma_{\text{rel}}$ measure the augmented training outperforms both, the ST100
and the ST20 benchmarks, cf. Fig.~\ref{fig:ggtt_mc_err}. In particular, it reaches the respective
peak performances after training with significantly fewer four-gluon events. 

Finally, we discuss the impact of the augmented training for the $gg\to t\bar{t} + 5g$ 
channel, which, to the best of our knowledge, has not been shown in ML-based phase-space sampling
works before. We begin by noting that the $\epsilon_{1\%}$ curve in the five-gluon case in
Fig.~\ref{fig:ggtt_1_percent_uw_eff} shows two prominent drops in performance in the ST100 training.
In fact, such drops can be observed for all training methods, especially in later iterations.
In iterations in which these performance drops occur, the models generate a few outlier weights.
As only a few events are affected, the performance metrics typically recover quickly. However,
to mitigate these performance dips, outlier weights might be clipped at a threshold value. 

When comparing $\epsilon_{1\%}$ of the first generated dataset, 
the informed-augmentation training yields an efficiency around 32 % 32.02
times higher than ST100, even though no actual $5g$ events were trained on, while ST100 has
seen around 42M actual $5g$ events. % 42,088,910
In comparison to ST20, this amounts to a factor around 68. This impressively
illustrates that the quality of the data used in the initial training steps outweighs sheer data
quantity. Similar to the other five-gluon process, the augmented training converges significantly
faster and to higher values than the two benchmark trainings. For $\epsilon_{1\%}$ the
informed-augmentation training plateaus around 100M training events\footnote{Performance figures quoted
  at exactly 100M, 200M, and 300M training events are obtained by linear interpolation using one
  reference point just before and after the relevant mark.}. 
Continuing training on another 100M events leads to a relative improvement of $\epsilon_{1\%}$ by
4\% only, cf. Fig.~\ref{fig:ggtt_1_percent_uw_eff}. % 4.26
In contrast, for ST100 and ST20 this leads to relative improvements by 49\% and 37\%, respectively, % 49.1,37.25
indicating their trainings have not yet reached a plateau. Accordingly, the informed-augmentation training 
can be stopped much earlier than the reference models, significantly reducing the need for evaluations of
the expensive $gg\to t\bar{t}+5g$ matrix elements, still providing significantly improved performance with
respect to the two benchmarks. Similar observations can be made when comparing performance saturations in
$\sigma_{\text{rel}}$, see Fig.~\ref{fig:ggtt_mc_err}. Again, the informed-augmentation training converges faster,
i.e.\ using fewer training events, achieving a better $\sigma_{\text{rel}}$ than the benchmarks. 

\paragraph{Practical considerations}
Lastly, we outline a practical workflow for applying the augmentation method. 
When building the training stack, one does \emph{not} need to start at the lowest 
possible particle multiplicity or phase-space dimension. For example, in our work, 
since the augmentation training gains most in the high-multiplicity learning problems, 
it suffices to train an $X+3g$ sampler to a reasonable performance, then start the augmentation 
to $X+4g$, followed by $X+5g$. A direct augmentation from $X+3g$ to $X+5g$ may also be 
possible. 

After establishing a mapping of condition labels in $\mathrm{\Phi}_{N+1}$ onto condition labels
in $\mathrm{\Phi}_{N}$, and identifying the dimensions in $\mathrm{\Phi}_{N}$ to copy and/or augment
to in $\mathrm{\Phi}_{N+1}$, a stacked training can be built up. Given the optimal amount of training data 
depends on the specific process and on available compute resources, a universal \emph{training}
prescription is not feasible. Instead, we recommend monitoring the training 
performance on the fly to decide after each iteration whether to continue or stop. 
To keep the optimisation stable, we recommend to clip large event weights. Furthermore, 
we recommend to compute a set of performance metrics on each iteration's generated dataset. 
These metrics can be combined to form an early-stopping criterion that halts 
training once further improvements become negligible, thereby avoiding unnecessary
integrand evaluations. 

\clearpage
\section{Conclusions}
\label{sec:conclusions}
We have introduced a simple yet effective training strategy that significantly 
reduces the cost of training phase-space point generators for higher-multiplicity 
final states, by cutting down on the number of matrix-element evaluations needed 
during training. The proposed method utilises models that have already learned a lower-dimensional 
projection of the to-be-learned phase space. In addition to being more efficient, the 
approach also yields generators with superior performance, given a finite number of 
training iterations. This training method 
exploits the factorisation of phase spaces 
\begin{equation*}
  \diff x_a \diff x_b \diff\mathrm{\Phi}_{N+1} \sim [\diff x_a \diff x_b \diff\mathrm{\Phi}_{N}] \times
 \diff\mathrm{\Phi}_{1}\,,
\end{equation*}
and, as discussed in Sec.~\ref{sec:knowledge-transfer}, this requires a parameterisation with the nesting property of
Eq.~\eqref{eq:coord_nesting}.

Because the proposed training method only initialises a proposal distribution that 
more closely resembles the actual integrand, exact event weighting preserves 
unbiased Monte Carlo estimates of physical observables and the generators reproduce the
correct target distribution.
The augmentation yields the greatest advantage when training a stack of multiplicities, 
as the training speed-ups accumulate. Moreover, it is most effective when applied to 
higher-dimensional phase spaces, since the larger the fraction of phase-space dimensions 
that can be inherited from a lower-dimensional model is, the larger the gain.

While here we concentrated on single partonic channels, physical jet-production processes
receive contributions from a number of partonic channels. In principle a dedicated sampler
can be trained for each individual channel, using augmentation to inform the training of
stacks of related channels. However, partonic channels can alternatively be grouped into
process classes that share a common integrator, see for example Ref.~\cite{Herrmann:2025nnz}.
In Ref.~\cite{Bothmann:2026yzp} a strategy for how to embed parton-flavour and helicity information
into Coupling-Layer Flows has been presented. We envisage that our here proposed augmentation
strategy can be used for the training of such conditioned flows in a straightforward manner.
All this requires in addition is a map between the flavour- and helicity labels of process
classes of different multiplicity that are mutually related.

This improved training technique is not limited to ODE-Flow models, but can be 
transferred to other adaptive sampling methods. Even classical algorithms such as \Vegas or 
Coupling-Layer-Flow models can benefit from the idea of reusing a well-sampled 
lower dimensional phase space. 
For \Vegas the transfer is particularly simple and requires no augmented dataset at all: the grid factorises over the
coordinates, so the $d_N$ one-dimensional grids inherited from the lower multiplicity can be copied directly, and the
three grids of the new particle initialised from those of an already-resolved one. Since \Vegas employs a single
remapping for all helicity configurations, the conditional augmentation of Sec.~\ref{subsec:conditional_augmentation} 
has no counterpart in this case. As for the flow, the copied grids only provide a starting point for the 
subsequent adaptation, which then has to account for the modified recoil kinematics and momentum fractions.
Moreover, the method is applicable beyond sampling phase spaces for stacks of 
increasing jet multiplicity. For example, when integrating a cross section at NLO accuracy, including
virtual and real-emission contributions, i.e.\

\begin{align}
  \sigma^{\text{NLO}} =& \int\diff \Phi_N[B(\Phi_N) + V(\Phi_N)]\nonumber\\ &+ \int \diff\Phi_{N+1}R(\Phi_{N+1})\,,\nonumber
  \end{align}

  one could augment a precisely learned Born phase space to account for the real-radiation
degrees of freedom and thereby improve the training of the real-emission component.

In this way the augmentation framework offers a versatile, simple, and cost-effective
tool for a broad range of importance-sampling applications in Monte Carlo integration.

\backmatter

%\bmhead{Supplementary information}
%
%If your article has accompanying supplementary file/s please state so here. 
%
%Authors reporting data from electrophoretic gels and blots should supply the full unprocessed scans for key as part of their Supplementary information. This may be requested by the editorial team/s if it is missing.
%
%Please refer to Journal-level guidance for any specific requirements.

\bmhead{Acknowledgements}
We are grateful for financial support from the German Federal Ministry of Research, Technology and Space (projects 05D23MG1 and 05H24MGA). 
The authors gratefully acknowledge the computing time granted by the Resource Allocation Board and provided on the supercomputer Emmy/Grete at NHR-Nord@Göttingen as part of the NHR infrastructure. The calculations for this research were conducted with computing resources under the project nip00094.
\onecolumn

\section*{Declarations}
The authors used OpenAI's ChatGPT OSS 120B, for more details see Ref.~\cite{2025arXiv250810925O}, to assist with writing and improving the 
readability of this manuscript. All generated suggestions were reviewed by the authors. The tool was not used to draw scientific conclusions. 
The authors take full responsibility for the final content. 
%%% Some journals require declarations to be submitted in a standardised format. Please check the Instructions for Authors of the journal to which you are submitting to see if you need to complete this section. If yes, your manuscript must contain the following sections under the heading `Declarations':
%%% 
%%% \begin{itemize}
%%% \item Funding
%%% \item Conflict of interest/Competing interests (check journal-specific guidelines for which heading to use)
%%% \item Ethics approval and consent to participate
%%% \item Consent for publication
%%% \item Data availability 
%%% \item Materials availability
%%% \item Code availability 
%%% \item Author contribution
%%% \end{itemize}
%%% 
%%% \noindent
%%% If any of the sections are not relevant to your manuscript, please include the heading and write `Not applicable' for that section. 

%%%\bigskip
%%%\begin{flushleft}%
%%%Editorial Policies for:
%%%
%%%\bigskip\noindent
%%%Springer journals and proceedings: \url{https://www.springer.com/gp/editorial-policies}
%%%
%%%\bigskip\noindent
%%%Nature Portfolio journals: \url{https://www.nature.com/nature-research/editorial-policies}
%%%
%%%\bigskip\noindent
%%%\textit{Scientific Reports}: \url{https://www.nature.com/srep/journal-policies/editorial-policies}
%%%
%%%\bigskip\noindent
%%%BMC journals: \url{https://www.biomedcentral.com/getpublished/editorial-policies}
%%%\end{flushleft}
\begin{appendices}

\section{Training-Method Validation}
\label{apx:training_validation}
This appendix serves as a validation for each training method's final model, by reporting all figures of merit used in the study. 
We present the phase-space efficiency, relative Monte Carlo integration error, $\sigma_{\text{rel}} = \frac{\sigma_{\text{error}}}{\sigma}$, Kish effective sample 
size, cf. Eq.~\eqref{eq:kish_ess}, and the 0.1\% and 1\% overweight unweighting efficiencies, see Eq.~\eqref{eq:uw_efficiency}. Final model evaluations 
are done for the $d\bar{d} \to e^+ e^- + 5g$, $gg \to t\bar{t} + 4g$ processes, in analogy to Ref.~\cite{Bothmann:2025lwg}, and for the newly introduced  
$gg \to t\bar{t} + 5g$ processes.
\begin{table*}[htpb]
    \centering
    \resizebox{\textwidth}{!}{%
        \begin{tabular}{lllllll}
            \hline
            Process & Training          & Phase-space eff. [\%] & $\sigma_{\text{rel}}$ [\%] & $N_{\text{eff}} / N_{\text{ev}}$ [\%] & $\epsilon_{1\%}$ [\%] & $\epsilon_{0.1\%}$ [\%]\\ \toprule
            \multirow{4}{*}{$d\bar{d} \to e^+ e^- + 5g$}
                & Stack unif.           & 75.68(1) & 0.043(1) & 48(1) & 8.32(4) & 3.2(1) \\ \cmidrule(lr){2-7}
                & Stack cp.             & 75.23(2) & 0.04192(6) & 48.7(1) & 8.55(4) & 3.55(6) \\ \cmidrule(lr){2-7}
                & Standard Training     & 75.44(1) & 0.050(8) & 39(5) & 6.68(5) & 2.6(2) \\ \cmidrule(lr){2-7}
                & Standard Training 20M & 75.81(1) & 0.05(1) & 37(6) & 6.02(7) & 2.3(2) \\ \toprule \toprule
            \multirow{4}{*}{$gg \to t\bar{t} + 4g$}
                & Stack unif.           & 88.50(1) & 0.0238(2) & 72.3(4) & 18.85(5) & 8.1(2) \\ \cmidrule(lr){2-7}
                & Stack cp.             & 88.25(1) & 0.026(3) & 68(4) & 17.96(6) & 8.2(3) \\ \cmidrule(lr){2-7}
                & Standard Training     & 87.762(4) & 0.0266(7) & 67(1) & 16.23(4) & 7.8(1) \\ \cmidrule(lr){2-7}
                & Standard Training 20M & 87.8412(98) & 0.0263(1) & 67.4(2) & 16.15(4) & 7.85(7)\\ \toprule
            \multirow{4}{*}{$gg \to t\bar{t} + 5g$}
                & Stack unif.           & 81.82(1) & 0.041(8) & 48(8) & 9.8(1) & 3.2(4) \\ \cmidrule(lr){2-7}
                & Stack cp.             & 81.76(1) & 0.038(3) & 51(3) & 9.82(5) & 3.5(3) \\ \cmidrule(lr){2-7}
                & Standard Training     & 82.850(9) & 0.0396(3) & 47.1(2) & 7.72(4) & 2.90(10)\\ \cmidrule(lr){2-7}
                & Standard Training 20M & 83.06(1) & 0.0409(6) & 45.2(6) & 7.28(6) & 2.8(2)\\ \hline
        \end{tabular}
    }
    \caption{Model performance after the final training iteration, evaluated on ten independent datasets. The phase-space
    efficiency and the relative Monte Carlo integration error are computed from the first 10M generated events of each dataset,
    including those with zero weight; the effective sample size and the unweighting efficiencies are computed from 10M
    non-zero-weight events. Quoted uncertainties are standard deviations across the ten runs, rounded to the first
    significant digit of the error.
    }
    \label{tab:performance_5g}
\end{table*}
\newpage
\section{Model Hyperparameters}
\label{apx:network-hyperparams}
In Tab.~\ref{tab:model-building-blocks} we collate the adjustable hyperparameters for the building blocks of our
CNF sampler. In Tab.~\ref{tab:trainable-params} we quote the resulting total number of trainable parameters for the
processes in the stacks $d\bar{d}\to e^+e^-+n g$ and $gg\to t\bar{t}+n g$. 

\begin{table}[htpb]
    \centering
    \begin{minipage}{0.48\textwidth}
        \centering
        \resizebox{\textwidth}{!}{%
        \begin{tabular}{lll}
            \toprule
            Hyperparameter  & Comment                                       & No. of                            \\
                            &                                               & parameters                        \\ \midrule
            Helicity        & No. of hel.                                   & $n_{\text{hel}}$-1                \\
            weight          & states: $n_{\text{hel}}$                      &                                   \\ \midrule
            Helicity        & embedding vector                              & $n_{\text{hel}}^2$                \\
            Embedding       & size: $n_{\text{hel}}$                        &                                   \\ \midrule
            $d_{\text{model}}$& 512                                           &                                   \\ \midrule
            Fourier feat.   & For sin/cos time encoding,                      &                                   \\
                            & embedding dimension:                                  &                                 \\
                            & $t_{\text{Fourier}} = 16 + 16$     & 16                                \\ \midrule
            Input layer     & $n_{\text{input}} =$                          & $n_{\text{input}} \cdot 512 + 512$\\
                            & $d_{N} + n_{\text{hel}} + t_{\text{Fourier}}$   &                                   \\ \midrule 
            Main MLP        & 4 layers à $512 \to 512$                      & $4\cdot(512\cdot512 + 512)$       \\ \midrule
            Output layer    & $512 \to  d_{N}$                               & $512 \cdot d_{N} + d_{N}$             \\ \hline
        \end{tabular}
        }
        \caption{Model building blocks and their number of adjustable parameters.}
        \label{tab:model-building-blocks}
    \end{minipage}
    \hfill
    \begin{minipage}{0.48\textwidth}
        \centering
        \begin{tabular}{lllll}
        \toprule
        Process & $n$ & $d_{N}$ & $n_{\mathrm{hel}}$ & No. model \\
                &     &     &                     & parameters \\
        \midrule
        \multirow{5}{*}{$d\bar d \to e^+e^- + ng$}
        & 1 & 7  & 8   & 1,078,878 \\
        & 2 & 10 & 16  & 1,086,249 \\
        & 3 & 13 & 32  & 1,098,300 \\
        & 4 & 16 & 64  & 1,120,863 \\
        & 5 & 19 & 128 & 1,169,058 \\
        \midrule
        \multirow{5}{*}{$gg \to t\bar t + ng$}
        & 1 & 7  & 32  & 1,092,150 \\
        & 2 & 10 & 64  & 1,114,713 \\
        & 3 & 13 & 128 & 1,162,908 \\
        & 4 & 16 & 256 & 1,280,799 \\
        & 5 & 19 & 512 & 1,611,810\\
        \bottomrule
    \end{tabular}
    \caption{Number of adjustable model parameters for the considered processes.}
    \label{tab:trainable-params}
    \end{minipage}
\end{table}
\vspace*{-0.8cm}
\section{Additional Plots}
\label{apx:additional_plots}
\begin{figure}[htpb]
    \centering
    \includegraphics[width=0.89\textwidth]{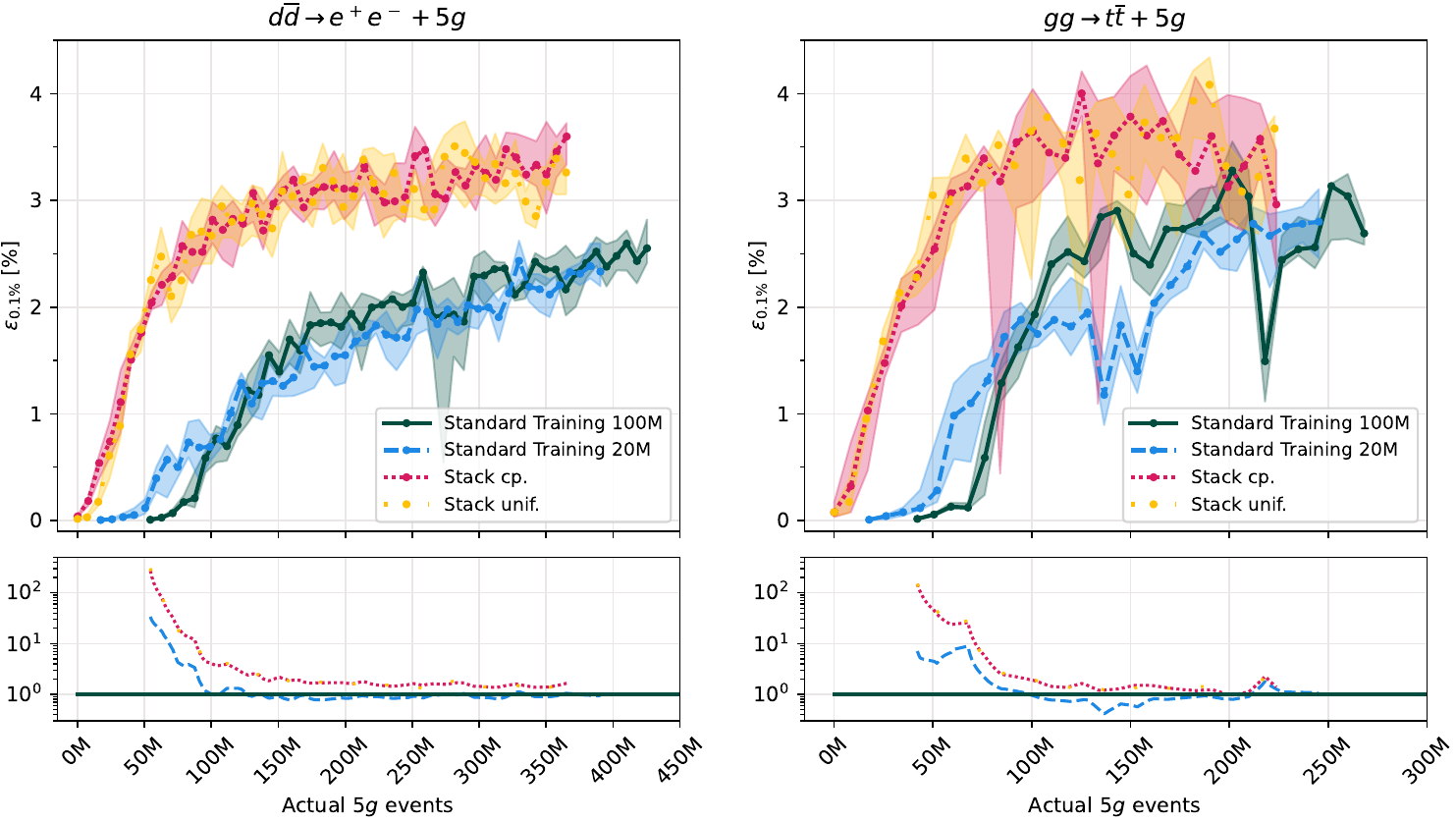}
    \caption{$0.1$\% overweight unweighting efficiency, cf. Eq.~\eqref{eq:uw_efficiency}, for the 
    $d\bar{d} \to e^+e^-+ 5 g$ and $gg \to t\bar{t} + 5 g$ processes, 
    comparing the standard training benchmark, standard training benchmark with $\sim20$M initial 
    uniformly sampled points, and the informed and uninformed augmentation training.}
    \label{fig:ddz5g_ggtt5g_01percent_uw_eff}
\end{figure}

\begin{figure*}[!t]
    \centering
    \includegraphics[width=0.89\textwidth]{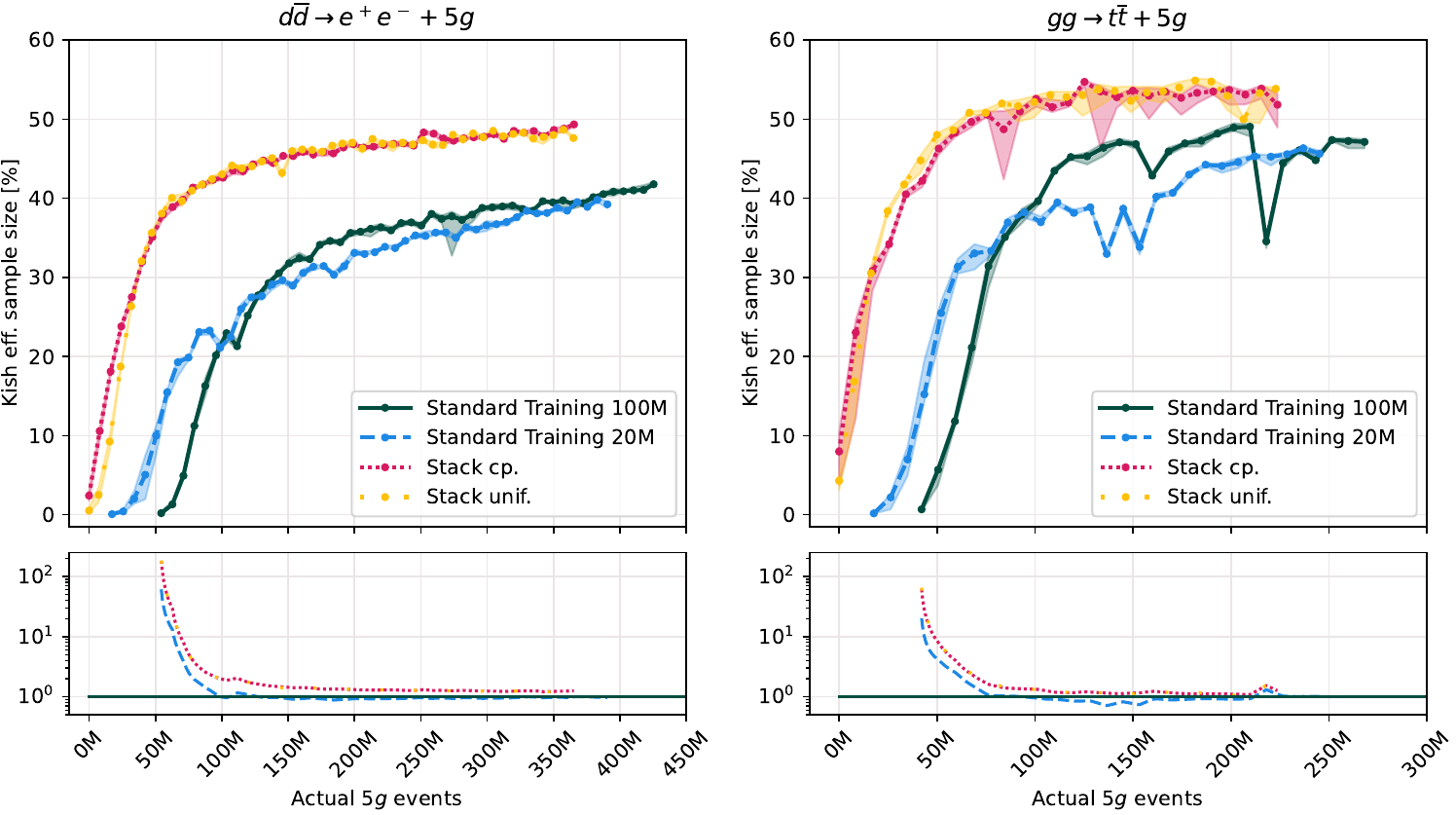}
    \caption{Kish effective sample size, cf. Eq.~\eqref{eq:kish_ess}, for the 
    $d\bar{d} \to e^+e^-+ 5 g$ and $gg \to t\bar{t} + 5 g$ processes, 
    comparing the standard training benchmark, standard training benchmark with $\sim20$M initial 
    uniformly sampled points, and the informed and uninformed augmentation training.}
    \label{fig:ddz5g_ggtt5g_ess}
\end{figure*}

In this appendix we show results for the $\epsilon_{0.1\%}$, cf. Fig.~\ref{fig:ddz5g_ggtt5g_01percent_uw_eff}, and Kish effective sample size,
cf. Fig.~\ref{fig:ddz5g_ggtt5g_ess}, figures of merit for the highest-multiplicity channels, i.e.\
$d\bar{d} \to e^+e^-+ 5 g$ and $gg \to t\bar{t} + 5 g$.
\twocolumn

\end{appendices}


\bibliography{main}% common bib file

\end{document}